\documentclass[conference]{IEEEtran}

\usepackage[available,functional,reproduced]{ndssbadges}
\usepackage[table]{xcolor}
\usepackage[inline,shortlabels]{enumitem}
\usepackage{algpseudocode}
\usepackage{graphicx} 
\usepackage[framemethod=tikz]{mdframed}
\usepackage{amssymb}  %

\usepackage{hyphenat}
\usepackage{multirow}

\usepackage[normalem]{ulem}

\usepackage{listings}
\usepackage[hyphens]{url}
\usepackage{hyperref}
\hypersetup{              
  colorlinks=true,
    linkcolor={blue!70!black},
      citecolor={blue!70!black},
      urlcolor={black}
  }

\usepackage{tcolorbox}
\lstdefinestyle{base}{
  moredelim=**[is][\color{red}]{@}{@},
moredelim=**[is][\color{blue}]{^}{^},
}

\newcommand\xoutpars[1]{\let\helpcmd\xout\parhelp#1\par\relax\relax}
\newcommand\soutpars[1]{\let\helpcmd\sout\parhelp#1\par\relax\relax}

\newcommand{\name}{Duumviri}
\newcommand{\el}{filter lists}

\newcommand{\BD}{Breakage Detector}
\newcommand{\bd}{breakage detector}
\newcommand{\Bd}{Breakage detector}
\newcommand{\TD}{Tracking Detector}
\newcommand{\td}{tracking detector}
\newcommand{\Td}{Tracking detector}
\newcommand{\subr}{partial-request}

\title{\name{}: Detecting Trackers and Mixed \\  Trackers with a  \BD{}}

\author{\IEEEauthorblockN{He Shuang}
	\IEEEauthorblockA{University of Toronto\\
		he.shuang@mail.utoronto.ca}
	\and
	\IEEEauthorblockN{Lianying Zhao}
	\IEEEauthorblockA{Carleton University\\
		lianying.zhao@carleton.ca}
	\and
	\IEEEauthorblockN{David Lie}
	\IEEEauthorblockA{University of Toronto\\
		david.lie@utoronto.ca}}

\newcommand{\subhead}[1]{\noindent{{\textbf{#1.} }}}

\def\showcomments{0}
\ifnum\showcomments=1
    \usepackage{todonotes}
    \newcommand{\fixme}[1]{{\textcolor{red}{[FIXME: #1]}}}
    \newcommand{\checkme}[1]{{\textcolor{orange}{[CHECKME: #1]}}}
    \newcommand{\ssnote}[1]{{\textcolor{blue}{[SS: #1]}}}    
    \newcounter{mynote}[section]
    
    \newcommand{\thenote}{\thesection.\arabic{mynote}}
    \newcommand{\viau}[1]{
         \textcolor{teal}{Viau: #1}}
    \newcommand{\dlnote}[1]{
         \textcolor{magenta}{david: #1}}
\else
    \newcommand{\fixme}[1]{}
    \newcommand{\checkme}[1]{}
    \newcommand{\thenote}[1]{}
    \newcommand{\ssnote}[1]{}
    \newcommand{\dlnote}[1]{}
    \newcommand{\viau}[1]{}
    \newcommand{\mdframe}
\fi

\begin{document}

\IEEEoverridecommandlockouts
\makeatletter\def\@IEEEpubidpullup{6.5\baselineskip}\makeatother
\IEEEpubid{\parbox{\columnwidth}{
		Network and Distributed System Security (NDSS) Symposium 2025\\
		23 - 28 February 2025, San Diego, CA, USA\\
		ISBN 979-8-9894372-8-3\\
		https://dx.doi.org/10.14722/ndss.2025.23267\\
		www.ndss-symposium.org
}
\hspace{\columnsep}\makebox[\columnwidth]{}}

\maketitle

\begin{abstract}

Web tracking harms user privacy.  As a result, the use of tracker detection and blocking tools is a common practice among Internet users. However, no such tool can be perfect, and thus there is a trade-off between avoiding breakage (caused by unintentionally blocking some required functionality) and neglecting to block some trackers. State-of-the-art tools usually rely on user reports and developer effort to detect breakages, which can be broadly categorized into two causes: 1) misidentifying non-trackers as trackers, and 2) blocking mixed trackers which blend tracking with functional components.

We propose incorporating a machine learning-based breakage detector into the tracker detection pipeline to automatically avoid misidentification of functional resources. For both tracker detection and breakage detection, we propose using differential features that can more clearly elucidate the differences caused by blocking a request. We designed and implemented a prototype of our proposed approach, \name{}, for non-mixed trackers. We then adopt it to automatically identify mixed trackers, drawing differential features at \subr{} granularity.

In the case of non-mixed trackers, evaluating \name{} on 15K pages shows its ability to replicate the labels of human-generated filter lists, EasyPrivacy, with an accuracy of 97.44\%. Through a manual analysis, we find that \name{} can identify previously unreported trackers and its \bd{} can identify overly strict EasyPrivacy rules that cause breakage. 
In the case of mixed trackers, \name{} is the first automated mixed tracker detector, and achieves a lower bound  accuracy of 74.19\%. \name{} has enabled us to detect and confirm 22 previously unreported unique trackers and 26 unique mixed trackers.
\end{abstract}

\section{Introduction}

Users navigating the web are constantly being monitored. 95\% of 21 million pages contain 3rd-party requests to potential trackers~\cite{10.1145/2872427.2883028}.  This extensive tracking results in a significant loss of privacy, as it allows users' sensitive information to be used for targeted advertising~\cite{liu2013adreveal}, behavioral profiling, and sold to third parties without their consent~\cite{olejnik2013selling}.  Therefore, there is a need to identify and block trackers to protect users' privacy. 

Web trackers come in two types: non-mixed trackers and mixed trackers. 
Non-mixed trackers send network requests that are purely for the purpose of tracking users. These requests may load tracking code onto the web client or send identifiers and information that enable users to be tracked. 
Since non-mixed tracker requests only contain tracker data and functionality, they are relatively easy to  identify and block. 
To make the identification and blocking of trackers harder, trackers can be mixed, meaning that requests made by tracking code contain both tracking and legitimate functionality~\cite{chen2021detecting, rack2023jack}. Most previous work has focused on non-mixed trackers, and has used on network request features~\cite{10.1145/2872427.2883028, gugelmann2015automated}, as well as both static~\cite{ikram2017towards} and dynamic JavaScript features~\cite{iqbal2020adgraph, siby2021webgraph} to identify and block such trackers. More recently, there have also been research proposals to deal with mixed trackers by attempting to identify mixed JavasScripts~\cite{castell2023astrack} and block the tracking functionality by disabling the tracking components~\cite{smith2021sugarcoat}.

In general, an effective tracker detector should have two objectives: 1) maximizing privacy by blocking as many trackers as possible and 2) minimizing web page breakage. Breakage can occur due to 1) incorrectly classifying non-tracking components as trackers and blocking them and 2) blocking mixed trackers in their entirety, which ends up blocking their functional component. 

Previous approaches train a single model to detect tracking without explicitly addressing the issue of breakage. This approach has two drawbacks:
1) The single tracking model must be highly accurate to avoid misidentification. Blocking a misidentified functional request can result in web page breakage. This is quite common---previous single-model approaches~\cite{iqbal2020adgraph} can cause breakage on 15\% of the sites. 
2) They are unable to block mixed trackers as they classify and block the request in its entirety, and thus are not able to block only the tracking component.

We explore adding explicit breakage detection into tracker detection. We argue that by explicitly considering the breakage introduced by tracker detection, we can minimize the detection imprecision 1) due to misidentification, where the detector made a false prediction and 2) due to mixed trackers, where the detector made a proper prediction, but still breaks a page.

This paper introduces \name{}, which incorporates two novel mechanisms to address both non-mixed and mixed trackers. First, as implied by its name\footnote{Duumviri is Latin for "two men," denoting a pair of officials sharing power and duty in ancient Rome.}, \name{} introduces breakage detection into tracker detection pipeline. It uses two models instead of one: one for detecting trackers, and the other for detecting which requests that, if blocked, will lead to web page breakage. This breakage detector can thus detect functional requests that were misclassified as tracking for non-mixed trackers, as well as request that contain both functional and tracking functionality for mixed trackers. \name{}'s models work on \textit{differential features}, which are derived from experimentally blocking requests and comparing the resulting page behavior to that of the original page. Using differential not only enables accurate detection of breakage but also enables \name{} to detect tracking request fields at \subr{} granularity, enabling \name{} to block tracking functionality without blocking legitimate functionality.

In designing \name{}'s \bd{}, we overcame the following challenges:
\begin{enumerate*}[nolistsep, leftmargin=*]    
    \item Feature selection. Previous proposals~\cite{blocked_or_borken} lacked the features for accurate breakage detection. We built our features by comprehensively covering the channels of externally-visible events emitted by a web page during rendering, effectively addressing the symptoms of web page breakages~\cite{nisenoff2023defining}.

    \item The lack of a dataset of breakages. It is challenging to find relevant breakage samples on live sites. We solved this by leveraging exception rules in filter lists for up-to-date collection of breakage samples that we reconstructed by ``flipping'' exception rules into block rules. 
\end{enumerate*}

\noindent This paper makes the following contributions:
\begin{itemize}[nolistsep, leftmargin=*]
    \item The \textit{design and implementation} of \name{}. The introduction of the \bd{} flags the breakage caused by misidentification and blocking mixed trackers, increasing overall accuracy. 
    We designed the detectors to use differential features enabling the blocking of request fields to block mixed trackers without causing breakage. 
    We trained the \bd{} by collecting reconstructable breakage samples and the \td{} for mixed trackers by collecting mixed request trackers from advanced content blockers such as AdGuard and UbO (Ublock Origin). 
    
    \item An \textit{evaluation of \name{} on non-mixed trackers identification} on 15K pages. Our results show that \name{} can reproduce the labels from human-generated \el{} with a 97.44\% accuracy. 
    Through our manual analysis of the disagreements between \name{} and the \el{}, we found 55\% of instances to be previously unreported new trackers. 
    In addition, 10\% of analyzed cases are filter list-caused web page breakages that \name{}'s \bd{} found. We have reported 22 cases of confirmed new trackers with 175 occurrences in our dataset and 2 instances of confirmed filter list-caused breakage to the community. 

    \item An \textit{evaluation of \name{} on mixed trackers identification}. We evaluated \name{} on all resources that \el{} deemed as non-trackers. Through a manual analysis, we found that \name{} can achieve a lower bound accuracy of 74.19\% in detecting tracking fields. In this process, we  found  26 mixed trackers with 83 occurrences in our evaluation dataset, which we have reported to the community. 
\end{itemize}

\subhead{Artifacts} Our artifacts are available on GitHub~\footnote{\url{https://github.com/dlgroupuoft/Duumviri-NDSS25}} and further discussed in Appendix \S \ref{appendix:artifact}.

\section{Background and Definitions} \label{sec:2}

We give  background information on web trackers and web page breakages.

\subhead{Web Trackers} We adopt a generic definition of web tracking, which is the process of (re-)identifying users in different computation contexts~\cite{snyder2023pool}. 
The computation contexts can be, and are not limited to, user-specific behavior, same-domain pages, browsing modes (e.g., incognito and regular), sites (i.e., cross-site tracking) and devices (i.e., cross-device tracking).  The tracking process involves, at a minimum, two stages: 1) storing a server-known identifier on the client or generating an identifier in the computation context and 2) retrieving the identifier or generating the same identifier in a different context.   

These two stages of tracking naturally lead to two types of tracking network requests: 
1) \textit{incoming tracking responses} that initiate client-side tracking in one computation context. For instance, a response may contain a set-cookie header that stores a server-generated unique identifier for stateful tracking or JavaScript payload that generates fingerprints for stateless tracking.
2) \textit{outgoing tracking requests} that contain privacy-sensitive information (e.g., a user identifier) in a different computation context. Such requests inform the tracking server when a specific action is performed by the user. 

Blocking the network requests in either stage  (i.e., a tracking response or a tracking request) prevents tracking. Blocking a tracking response prevents storing identifiers or generating identifiers in one context; blocking a tracking request prevents a server from learning that a specific user in a different computation context.  

\subhead{Mixed Trackers} Broadly speaking, a mixed tracker is a tracker that also has legitimate functionality. As the tracking process has two stages, each stage leads to a type of mixed trackers:  
1) A mixed response tracker contains a response with mixed tracking and legitimate functionality. For instance, a response may contain JavaScript code that handles web page interaction yet also dynamically generates user fingerprints~\cite{rack2023jack}.
2) A mixed request tracker is a request with request fields (e.g., query string parameters, cookies) that have mixed  tracking and legitimate functionality. 
To illustrate, we provide a real-world example: after clicking on a search result on a popular search engine, the page sends one request (URL shown in Listing \ref{tab:mixed_tracker}) that redirects the user to the desired page. In the URL, query string parameters \textit{goods\_id} and \textit{sku\_id} are functional parameters that redirect to the specific product that the user clicked on, but \textit{\_x\_ns\_msclkid} is a tracking parameter (assigned by Bing) that records an ID of the click. 
This example mixed request tracker cannot be blocked in its entirety: doing so (or removing any of the functional parameters) breaks redirection as the request will land on a generic page as opposed to the product the user intends to see. However, removing the only tracking parameter while keeping the others stops tracking while preserving the redirection.

\begin{lstlisting}[style=base, caption={An example mixed request tracker. Blue indicates functional parameters. Red indicates tracking parameters. Additional parameters omitted for brevity.}, captionpos=b, basicstyle=\footnotesize, label=tab:mixed_tracker, breaklines=true, float=!t, belowskip=-1 \baselineskip, ]

https://www.temu.com/subject/n9/googleshopping-landingpage-a-psurl.html?^goods_id^=601099526089385&^sku_id^=17592258865022&@ _x_ns_msclkid@=eeec99c83e911b00583ffc4bc3e34060 
\end{lstlisting}

\begin{figure}
    \centering \small

    \includegraphics[trim={0 0 0.8cm 0},clip,width=\linewidth]{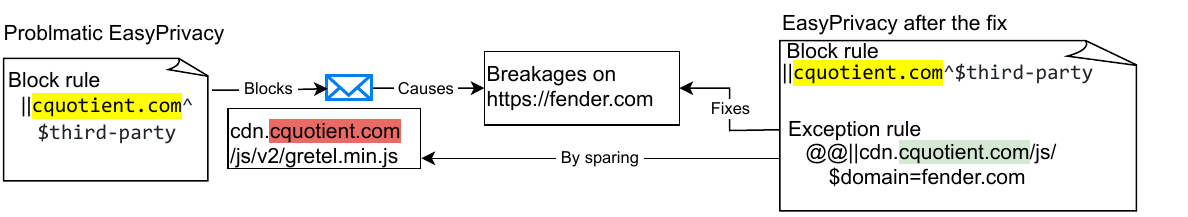}
    
    \caption{An example of an exception rule used to `fix' page breakage. When `fender.com' fetches `gretel.min.js' from `cdn.cquotient.com'. This request is blocked as the domain is listed as a tracking server. However, the particular resource is used for legitimate web page functionality (product recommendation); blocking it causes missing page content. Privacy developers fix this issue by adding an exception rule that makes an exception for `fender.com'~\cite{github_commit_fixing}.}
    \label{fig:exception_rule}
    \vspace{-3mm}
\end{figure}

\noindent \textbf{Web Page Breakage} can occur when 1) privacy developers make mistakes in addressing trackers and 2) erroneously blocking mixed trackers at the request granularity. 
Human mistakes can appear in any step of the tracker addressing work flow (e.g., incorrectly identified tracker, an incorrect fix, or failing to identify breakage). The chance of a mistake increases as the number of rules in the filter lists increases. 
Blocking mixed trackers entirely at the request granularity can lead to web page breakage since the functional request fields are also blocked.

Currently, when a breakage occurs and is experienced by a content blocker user, she reports it to the privacy developers who then address the breakage. While this process can be long and error-prone~\cite{snyder2018filters}, it is generally difficult for a privacy developer to manually perform breakage evaluation comprehensively due to the lack of domain knowledge on the broken site (e.g., unfamiliarity, site requiring an account for access, site is restricted to certain geographic locations). 
One common method for fixing breakages is by inserting a new exception rule. The exception rule spares the erroneously blocked content, where we show one example in Fig \ref{fig:exception_rule}. 
Such practices can lead to performance degradation due to the number of exception rules and maintainability of the filter lists. 
An automatic breakage detector can help tracker detection test candidate rules constructed during tracker identification by catching erroneously identified trackers automatically. %

\section{The Design of \name{}} \label{sec:3}

\begin{figure*}
    \centering \small
    \includegraphics[width=1\linewidth]{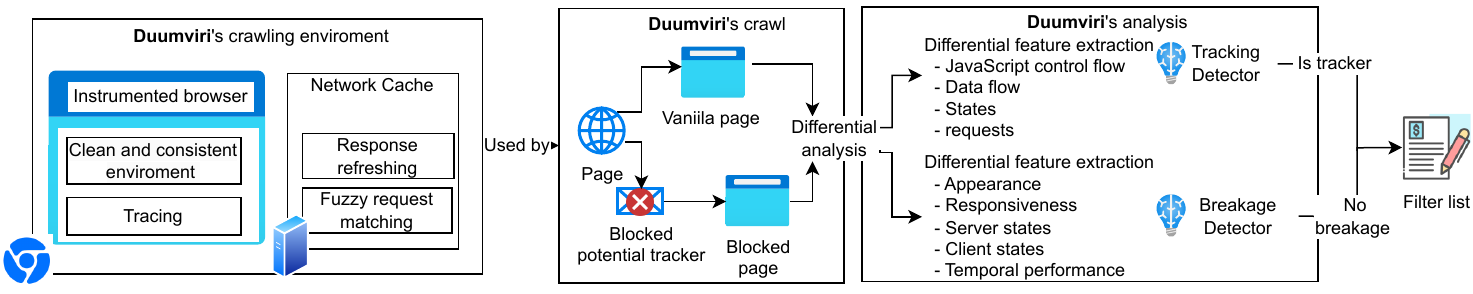}
    \caption{1) \name{} visits a page using two instrumented browser instances capable of produce a rendering trace. Both instances share a network cache. 2) \name{} conducts differential analysis on the page instances and draws differential features independently for its detectors. 3) the detectors take the features and make predictions. The predictions determine if the potential tracker is added to the filter list.}
    \vspace{-1mm}
    \label{fig:components}
\end{figure*}

We describe \name{}'s design and how it enables detection of non-mixed trackers. We begin with \name{}'s workflow, followed by \name{}'s features and finally, how \name{} collects its dataset for training detectors.

\subhead{Workflow} \name{}'s workflow mirrors the approach that one might imagine is taken by a human  privacy developer in addressing trackers. 
With components in Fig \ref{fig:components}, given a page-under-analysis ($PU\!A$), \name{} iteratively selects an outgoing request-under-analysis ($RU\!A$) and executes the following steps:
\begin{enumerate*}[nolistsep, leftmargin=*]    
    \item Differential Page Visits. \name{} visits the $PU\!A$ to gather a trace of the page rendering process, denoted $T$. This trace is used for differential feature extraction and is described in detail in \S \ref{sec:traces}. Next, \name{} revisits $PU\!A$ with $RU\!A$ blocked, generating another trace, $T_B$. Blocking a request means intercepting it and terminating it with a response of status code 403.

    \item Differential Feature Extraction. \name{} compares traces $T$ to $T_B$ producing two sets of differential features $F_{tracker}$ and $F_{breakage}$ for the \td{} and \bd{}, respectively. 

    \item Tracker Identification. \name{} invokes its \td{} with $F_{tracker}$ to obtain a label indicating whether $RU\!A$ is a tracker. If $RU\!A$ is not classified as a tracker, \name{} skips the next steps and proceeds to the next $RU\!A$.

    \item Breakage Evaluation. If $RU\!A$ is classified as a tracker \name{} invokes its \bd{} with $F_{breakage}$ to obtain a label indicating whether $RU\!A$ breaks the page when blocked.   

    \item Automatic Fixing. If the \bd{} does not detect breakage, then \name{} creates a block rule for future $RU\!A$. If the $RU\!A$ is deemed to be a tracker but also causes breakage when blocked, it is labeled a  potential mixed tracker, which is described in \S \ref{sec:adopting_for_mixed}. 
    
\end{enumerate*}

\subsection{Differential Features}
We designed differential features for \name{}'s \bd{} and \td{} because 1) for the \bd{}, differential features describe the \textit{change} in web page state, which contains more information than just looking at a single page. 2) for the \td{}, differential features provide accurate attribution of potential tracking activities to the $RU\!A$ and 3) we can draw differential features from blocking requests for non-mixed tracker analysis as well as from blocking request fields for mixed tracker analysis.  
Due to space constraints, we present the top 10 most important features of the detectors in Table \ref{tab:features_importance_both} along with a high-level categorization.

\begin{table*}
    \centering 
    \rowcolors{2}{white}{gray!15}
    \caption{Top-10 most important features of \name{}'s detectors. ``Importance'' shows the permutation feature importance using accuracy as the metric, repeated 5 times. ``Uniq'' illustrates whether the feature is proposed by and unique to \name{} compared to previous works in the area. }

    \begin{tabular}{p{5cm}cc    |p{6.7cm}cc}
    \rowcolor{gray!50}
    \multicolumn{3}{c|}{\BD{}} & \multicolumn{3}{c}{\TD{}} \\
     \rowcolor{gray!50}    Feature  & Importance & Uniq & Feature  & Importance & Uniq \\
        $\Delta$ console logs & 3.2 & $\checkmark$ &  $\Delta$ parameters of the blocked request & 14.8 &   \\
        $\Delta$ page load time & 2.81 & $\checkmark$ & $\Delta$ URL length of the blocked request & 8.08  &\\
         $\Delta$ event listeners & 2.25 & $\checkmark$ & $\Delta$ response size of the blocked request & 3.87 &\\
         $\Delta$ cookies values & 1.07 & $\checkmark$ &  $\Delta$ times first party appear in the blocked request  & 3.42 &\\
         $\Delta$ document height & 0.79 & $\checkmark$ & $\Delta$ `eval' appear in the response of the blocked request & 3.23 &\\
         $\Delta$ CSS classes  & 0.73 & $\checkmark$ &  $\Delta$ high entropy fingerprinting function executed & 0.54 &\\
         $\Delta$ DOM tree & 0.38 & $\checkmark$ & $\Delta$ third party requests blocked & 1.03 &\\
         $\Delta$ listeners on interactable elements& 0.35 & $\checkmark$ & $\Delta$ requests blocked & 0.3 &\\
         $\Delta$ HTML tag sequences  & 0.33 & & $\Delta$ third party requests with sensitive information & 0.29 & $\checkmark$\\
         $\Delta$ full-paged screenshot as a feature vector  & 0.31 & $\checkmark$ &  $\Delta$ `eval' in the ancestors nodes of the blocked request & 0.24 &\\
    \end{tabular}
    \vspace{-3mm}
    \label{tab:features_importance_both}
\end{table*}

\subhead{\BD{} Features} We designed \bd{} features by modeling web page breakage. We define web page breakage as changes in browser behavior in at least one externally visible channel compared to the vanilla functional page. This could include the absence of certain user interface (UI) elements affecting appearance or the lack of event listeners causing unresponsiveness. We consider the following externally visible channels: 1) web page appearance for user perception, 2) user input handling for interactiveness, 3) network requests for server-side states, 4) writes to persistent storage for client-side states, and 5) temporal performance for user experience. 

We developed a total of 63 differential features. Due to space constraints, we describe the complete features in Appendix. As shown in Table \ref{tab:features_importance_both}, 9 out of the 10 most important features are unique to \name{}. 
We further show that our \bd{}, which relies on these features, contributes to \name{}'s accuracy through an ablation analysis in \S \ref{sec:eval:ablation}.

\subhead{\TD{} Features}
We designed the differential features for the \td{} in four broad categories:
\begin{enumerate*}
\item DOM states capture the change in DOM elements and event listeners. Trackers often rely on tracking pixels to send sensitive information or event listeners to trigger request sending.
\item Requests capture communication between a tracker and the remote server, as trackers usually rely on network requests to share sensitive information like user identifiers.
\item JavaScript control flow captures unique tracking activities, such as invoking high-entropy APIs, which differ from functional scripts.
\item Data flow features capture information flow, such as cookies, from one actor (e.g., a network request) to local storage or other actors (e.g., scripts), which is crucial as much tracking information (e.g., user identifiers) must be shared with remote servers to complete tracking.
\end{enumerate*} 

In the rest of this section, we will detail the trace, followed by how differential features are constructed, and finally, we discuss how we ensure reliable features extraction (e.g., server-side randomness, session-specific requests). 

\subsubsection{Traces} \label{sec:traces}
The traces that \name{} collects are used for building differential features. Each trace contains the following components:
\begin{enumerate*}[nolistsep, leftmargin=*]
\item Requests: This component captures the direction, timing, and content (i.e., headers, body) of all network requests during page rendering. \name{} uses a man-in-the-middle proxy to intercept all requests in decrypted form, enabling access to request plaintext.
\item DOM elements: This component captures the raw DOM elements that are of interest to us. We currently track elements that we believe have a correlation to tracking or breakage, including canvas, audio, buttons, input, span, video, image, script, and hyperlink tags. For each element, we track the layout, position, and content.
\item Events: We track DevTool events such as when a page downloads (downloadWillBegin) and when a page finishes loading (loadComplete).
\item Event listeners: This component tracks all event listeners, including listener type, details (whether it is passive or fired once), the target object (e.g., a button), and the event handler (e.g., JavaScript text).
\item Scripts: This component tracks all parsed JavaScript by V8, including external and inline scripts. For both types, we track the script's textual content, position and the source URL if it is an external script.
\item Appearance: This component tracks information related to page appearance. We track a ``long'' screenshot covering all visual elements, the list of loaded fonts, the list of colors used on all HTML elements, the inner text, the main text of the page~\cite{barbaresi-2021-trafilatura}, and all CSS styles.
\item Storage values: This component describes the storage values including cookies, session, and local storage.
\item Console log: This component contains the console logs that a page prints. We track the level (e.g., serve, message), time, source, and the actual log message.
\item Perceptual adblocker: This component tracks the number of ads found by AdHighlighter~\cite{storey2017perceptual}.
\item PageGraph: This component is the complete PageGraph~\cite{PageGrap4:online}. PageGraph is a graph representation of the page rendering process, including nodes describing entities (e.g., HTML element, web resources, JavaScript files), and edges describing the actions (e.g., resource loading, DOM modifications) by the web entities.
\end{enumerate*}

\subsubsection{Differential Features Construction}
Given two traces $T$ and $T_B$, \name{} produces the differential features based on the type of trace component. 
A differential feature describes the degree of differences between two trace components. The goal of feature construction is to obtain accurate and consistent differences. \name{} has two default methods for extracting coarse-grained differences: hamming distance and cosine similarity. 
While they work well for features such as fonts and colors where we only need to know the degree of differences, they are not suitable for complex comparisons on components such as requests and event listeners. Below, we will describe the reasons and our solution: 
\begin{enumerate*}[nolistsep, leftmargin=*]
    \item Requests. Network requests can be session-specific (e.g., random values for cache busting~\cite{HowGoogl82:online}). However, we need to consider requests of the same origin as identical for accurate feature extraction. To do this, \name{} implements \textit{fuzzy request matching} based on request initiator, type (e.g., POST), URL, headers and body. Specifically, we consider two requests to be the same if the request initiator and type are identical, and the URL, headers and body have over 95\% similarity as measured by cosine similarity. 

    \item DOM elements. Similar to requests, the element may be session-specific (e.g., dynamically generated attributes). \name{} considers two DOM elements to be the same based on a combination of the structural and stylistic similarity as described in \cite{gowda2016clustering}.     
    \item Scripts. We use cosine similarity to measure the similarities among JavaScript text after vectoring the JavaScripts into token counts.  

    \item Event listeners. We consider two listeners the same if the type of the event, target object (a DOM element), and handler (a JavaScript function) are all identical.   
    
    \item Appearance. A common method for comparing screenshots is pixel-based similarity~\cite{castell2023astrack, nikiforakis2015privaricator}. However, we found this method to be unreliable for two reasons: 1) benign changes may occur due to the non-deterministic behavior of web pages (which we detail below), and 2) not all visual changes equally contribute to breakage. For example, a slight change in the position of an image could result in a large percentage of pixels being different. Instead of directly comparing pixels, \name{} compares the feature vectors of the screenshots using a pre-trained EfficientNet model~\cite{tan2019efficientnet}. We chose a vision-based model because it mimics what a human user perceives. And, EfficientNet has demonstrated generalization capabilities that often extract semantically meaningful features. By comparing the feature vectors, we can assess ``how semantically similar the two screenshots are as perceived by a human user'', which, based on our experience, has a higher correlation to breakage than pure pixel-based similarity measurement. 

    \item PageGraph. We do not measure the similarity of page graphs at the graph granularity. We derive lower-level features as used in AdGraph~\cite{iqbal2020adgraph} and WebGraph~\cite{siby2021webgraph} and draw differential features using the default methods on those features, which we describe in Appendix. 
\end{enumerate*}

\subsubsection{Obtaining Reliable Features} \label{sec:design:fo:q2}
One issue in generating differential features is the non-deterministic behavior of web pages, which can arise from various factors involved in the web page rendering process, such as server-state (e.g., network responses), client-state (e.g., existing cookies), and characteristics (e.g., user agent), as well as the web page content itself (e.g., time-sensitive content). This non-deterministic behavior introduces noise into the differential features that are not caused by blocking the request, increasing the likelihood of the \bd{} mistakenly predicting breakage. Such erroneous breakage predictions then decrease \name{}'s overall accuracy by incorrectly identifying trackers as non-trackers. To mitigate this impact on accuracy, \name{} employs two techniques: minimizing web page non-deterministic behavior through a specialized rendering environment and rendering each page a configurable number ($k$) times. 
While we provide details on our rendering environment below, rendering a single page $k$ times results in $k$ vanilla page traces and $k$ page traces of $T_B$ with a blocked request-under-analysis ($T_B$), totaling $k^2$ pair-wise differences. Given that non-deterministic behavior occurs less frequently than normal behavior, \name{} retains the most commonly occurring differential features out of all $k^2$ differences. To illustrate this technique with an example, when visiting $colgate.com$, a survey dialog may randomly appear, prompting for feedback. If this dialog appears during the differential feature extraction process, \name{} may incorrectly attribute differences in the user interface to the blocked $RU\!A$. However, since this dialog does not appear most of the time, by visiting the page multiple times, we obtain the majority of differences without the survey dialog. As a result, \name{} can accurately extract the differences caused by the $RU\!A$.

We explain \name{}'s specialized rendering environment that minimizes web page non-deterministic behavior as follows.
\begin{enumerate*}[nolistsep, leftmargin=*]
    \item Specialized Environment: Servers may react based on client states such as existing cookies, user agents (e.g., desktop, mobile), window sizes, and client locations (i.e., IP addresses). \name{} ensures a consistent, clean (e.g., no cookies), and plausible (like that used by human users) pre-loading environment.

    \item Network Cache: Some servers may send back non-deterministic responses, such as third-party ads. This non-determinism can affect page appearance.
    To deal with this, \name{} uses a network cache to record the communication (e.g., requests and responses) to the server when rendering the vanilla page and replays the same response when rendering the second page if an identical request passes through the cache.

    \item Dynaminism Mask: There may be non-deterministic HTML elements such as time-sensitive elements (e.g., slide shows, videos). We observed that the layout of the page is usually deterministic; we use the following method to address non-deterministic element appearance: for every screenshot, \name{} takes another screenshot after a configurable amount of time. The difference between the two screenshots forms a dynamism mask — these are the time-sensitive and non-deterministic areas. These areas are excluded from appearance comparison during feature extraction.

\end{enumerate*}
Still, \name{} cannot and does not aim to eliminate all non-deterministic behavior; it will benefit from continued  progress in this area~\cite{goel2022jawa}.

\subsection{Training Dataset} \label{sec3:training_dataset} \label{design:f_oracle:data}   \label{sec:design:fo:q3}

\begin{table}[t]
    \centering 
        \caption{The training size, test accuracy, and cross-validation (5 folds) accuracy of the detectors. }
    \label{tab:models_test_accuracy}
    \rowcolors{2}{gray!15}{white}
    \begin{tabular}{p{2cm}|c|cc|cc}
    \rowcolor{gray!50}
      & Training  &  \multicolumn{2}{c|}{Test} & \multicolumn{2}{c}{Cross-validation} \\
\rowcolor{gray!50} \multirow{-2}{*}{Detector}  & Size  & Size  & Accuracy  & F1  & STD \\
        Breakage  & 15,854  & 3,171 & 98.30   & 0.9591 & 0.0028 \\
        Tracker  & 27,721 & 5,545 & 93.62  & 0.9268 & 0.0026\\
        Mixed Tracker  & 1,976 & 396 & 85.10 & 0.8499 & 0.1923 \\
    \end{tabular}
\vspace{-3mm}
\end{table}

\subsubsection{\Bd{}} \label{sec:design:bd}
One problem in training the \bd{} is the lack of positive data points---i.e., samples of real breakage. While previous work reconstructs breakage from commit messages~\cite{blocked_or_borken}, our evaluation shows that it is an unreliable source (\S \ref{eval:breakge:q2}). Instead, \name{} gathers breakage samples from the exception rules in the current filter lists. While exception rules are used by privacy developers to temporarily ``fix'' broken pages (\S \ref{sec:2} and Fig \ref{fig:exception_rule}), they have two practical advantages 1) exception rules are constantly validated by filter list developers to ensure relevancy---irrelevant rules that no longer fix breakage are removed by developers and 2) exception rules are written and tested by developers to exclude the exact resource that causes breakage. We discuss these points in more detail in \S \ref{eval:breakage}.

\name{} reconstructs a breakage by \textit{flipping} an exception rule into a block rule. This process blocks the exact resource whose blocking causes breakage and it reconstructs the exact breakage that the privacy developer faced before adding the exception rule. Specifically, \name{}: 1) flips the exception rule to a block rule 2) navigates to the URL indicated by the domain modifier associated with the rule and 3) generates differential features by comparing the vanilla page and the resulting page. 

Not all exception rules can reconstruct the breakage. While developers do monitor these rules, some can still inevitably become stale (e.g., page has changed but the rule is not updated). In addition, the domain specifier can be ambiguous (e.g., the specifier points to a domain, but the breakage only occurs on a specific page within that domain). When reconstructing breakage, \name{} identifies such cases by monitoring the number of resources blocked by the flipped rule. Flipped exception rules that do not block any resources are discarded. We demonstrate the effectiveness of this detection in \S \ref{eval:breakge:q1}. We gathered a total of 13,921 exception rules and constructed 2,308 breakages, which were sampled to ensure correctness.

For the non-breakage samples, we need page changes that do not disrupt functionality. We want the \bd{} to ``accept'' legitimate changes. One of the legitimate change is the blocking of trackers. Thus, we draw differential features typical to blocking trackers using EasyList and EasyPrivacy as the ground truth. While these filter lists themselves may be inaccurate and cause breakage, we only collect non-breakage samples from the top 5K sites from the Alexa Top List, generated in May 2022. Popular sites tend to have issues discovered and resolved more quickly due to their large visitor numbers.

The \bd{} was trained on 15,854 data points, comprising 2,308 reconstructed breakages making up the positive label (quality evaluated in \S \ref{eval:breakge:q1}) and 13,546 cases of regular trackers as the negative label. We use an XGBoost~\cite{XGBoostD91:online} model.

\subsubsection{\Td{}} We train our detectors using requests encountered while crawling\footnote{We use the term `crawl' in this paper to mean visiting the URL and wait for page to load. } the top 5K sites from the Alexa Top 1M List generated in May 2022. Using EasyList and EasyPrivacy as ground truth, we have 12,936 (46.66\%) cases of trackers and 14,785 (53.34\%) cases of non-trackers. We also use XGBoost model.

\section{Adapting \name{} for Mixed Trackers} \label{sec:adopting_for_mixed}

In this section, we discuss how we adopt \name{} to automatically identify mixed trackers. Based on the definition of tracking from \S \ref{sec:2}, we now describe a baseline model of how users deal with mixed trackers. 
Filter lists such as EasyList and EasyPrivacy cannot deal with mixed trackers directly. They check resources at request granularity: a request is either blocked or spared entirely. Such a crude decision implies a lose-lose situation for mixed trackers: blocking a mixed tracker entirely potentially breaks the web page due to the blocking of the functional request fields; sparing a mixed tracker hurts user privacy. 
Advanced content blockers, such as UbO (Ublock Origin) and AdGuard, perform analysis at \subr{} granularity. A request field is a component of a HTTP request whose content is application-defined. Instances of request fields include query string parameters, request headers and body. 
UbO and AdGuard have added the ability to  inspect and alter individual request fields, enabling them to remove only the request fields of mixed trackers, thus preserving website functionality. We list the rules in UbO and AdGuard that are capable of addressing mixed trackers including the type of mixed trackers they address in the first two columns of Table \ref{tab:mixed_tracker_stats}.

\subhead{Prevalence of Mixed Trackers} \label{sec:mixed_tracker_prevalence} Using these rules, we conducted a prevalence study on both types of mixed trackers. First, we gathered filter lists from content blockers compatible with UbO and AdGuard's rule syntax, including UbO, AdGuard, Adfilt, and ClearURLs. All rules utilizing this syntax as of March 2024 were collated and applied to the traffic encountered during the crawl of 15K pages in \S \ref{sec:eval:q2}. We collected tuples of (domain, requests, and applied rules) for cases flagged by any rule and exclude requests that are blocked entirely by UbO or AdGuard, as these are likely non-mixed trackers. The results, presented in the last five columns of Table \ref{tab:mixed_tracker_stats}, reveal that the majority of mixed trackers fall under the mixed request type. It is worth noting that, as these rules can address a broader range of undesired information leakage within requests beyond trackers (e.g., performance measurement), the statistics collected may overestimate the real number of trackers. To reduce the overestimation, the filter lists we used are from privacy-focused content blockers.

\begin{table}[t]
    \centering    
        \caption{Statistics on currently identified mixed trackers on top 15K sites. \protect\footnotemark[1] a UbO rule. \protect\footnotemark[2] an AdGuard rule.}
    \label{tab:mixed_tracker_stats}
    \scriptsize
    \rowcolors{2}{gray!15}{white}
    \begin{tabular}{p{0.9cm}|c|c|ccc}
    \rowcolor{gray!50}
 Type   & Rule Type & \# Rules & \# Instances & \# Uniq  & \# Domains \\

     & redirect\footnotemark[1]\footnotemark[2] & 1,914 & 9,052 & 7,214 & 4,197\\
  \cellcolor{white} &  replace \footnotemark[2]  & 718 & 52 & 48 & 18 \\
 
           &  empty \footnotemark[1] & 5 & 0 & 0 & 0 \\
 \cellcolor{white} &  jsonprune \footnotemark[2] & 26 & 0 & 0 & 0 \\
          
\multirow{-5}{0.9cm}{Mixed Response}          &  hls \footnotemark[2] & 2 &0 &0 &0 \\
        
  & removeparam \footnotemark[1]\footnotemark[2] & 130 & 578 & 485 & 357\\
 
\cellcolor{gray!15}   & removeheader\footnotemark[2] & 2 & 0 & 0 & 0\\
 \cellcolor{gray!15} \multirow{-3}{0.9cm}{Mixed Request}  &  cookie\footnotemark[2] & 717 & 108,491 & 98,923 & 4,904\\
        
    \end{tabular}
    \vspace{-3mm}
\end{table}

In this work, we thus focus on blocking mixed trackers by blocking mixed requests---that is requests that contain both functional and tracker parameters. The reasoning for this is that 1) mixed trackers are prevalent  as shown by Table \ref{tab:mixed_tracker_stats}, and 2) blocking either one of mixed requests and mixed responses can equally prevent tracking by mixed trackers. The downside of this approach is that mixed response trackers still execute on the client machine, potentially wasting computation resources. We believe that \name{}'s method can be extended to identify mixed response trackers, but leave an exploration of this to future work.

In the current implementation, \name{} performs analysis on the request field by assuming that tracking information and functional information reside in different request fields.
This assumption is reasonable as making each request field either for tracking or other functionality eases server-side request field parsing, and the fact that there is no known tracker that mixes tracking and functional information within a single request field. In fact, there is a proposal to extend UbO to block  at sub-field granularity~\cite{Implemen63:online}, but it has not been implemented due to the lack of evidence of need for this feature. 
Nonetheless, we designed \name{} to be extensible to conduct analysis at partial-request field granularity,  provided that methods for separating tracking information from functional information within a mixed request field are available. 
For instance, assuming one POST request body is mixed as proposed in \cite{Implemen63:online}, one separation method is to use content type-specific parsing: using regular URL parsing for content type \textit{application/x-www-form-urlencoded} and JSON parsing for \textit{application/json}. We leave the implementation of this separation method as future work.

\subhead{Workflow} Given an $FU\!A$, \name{} detects whether it contains tracking information on a page-under-analysis ($PU\!A$) by performing the following steps. \begin{enumerate}[nolistsep, leftmargin=*]
    
    \item Differential Page Visits. \name{} renders the vanilla page once to produce a page trace ($T$). It then revisits the $PU\!A$, and intercepts the request containing $FU\!A$ and modifies the request by blocking $FU\!A$ from it. This modified request is sent to the server, whose response (from the server) is sent back to the page. This process produces another trace ($T_B$).
    
    \item Feature Extraction and Identification. \name{} extracts the differential features $F_{tracker}$ and $F_{breakage}$, from the traces $T$ and $T_B$. \name{} then classifies whether a particular $FU\!A$ is used for tracking or not by invoking \td{} with $F_{tracker}$. Similarly, it infers whether the $FU\!A$ breaks the page when blocked by invoking \bd{} with $F_{berakage}$. 
    
    \item Fixing. For all $FUAs$ that are used for tracking, \name{} automatically creates filter rules blocking them in UbO's syntax: $removeparam$ and $cookie$ syntax, preventing tracking. If a $FU\!A$ is both tracking and breaks the page when blocked, it means the $FU\!A$ contains both tracking and functional information not separated by request structure violating \name{}'s assumption. We leave such cases as non-tracker as \name{} cannot currently handle such a case. 
\end{enumerate} 

\subhead{Training Dataset} We trained a new \td{} for mixed trackers using the mixed trackers found using existing filter lists and tools in the prevalence study in \S \ref{sec:mixed_tracker_prevalence}, as this identified specific requests fields that privacy developers have labeled as being used for tracking. We use them as a ground truth for mixed tracking request fields. Although they may contain noise as discussed in \S \ref{sec:2}, they are the best mixed request tracker samples we can find. Specifically, we collected 71 and 905 cases of tracking parameters and cookies totaling 976 instances of tracking request fields. We randomly sampled 1,000 cases of non-tracking request fields from two sources: 1) functional parameters from the request not blocked by EasyPrivacy and 2) functional cookies as indicated by Cookiepedia~\cite{cookiepedia}. This process yields a balanced dataset of 1,976 data points in total. We again use an XGBoost model. We report the test accuracy in Table \ref{tab:models_test_accuracy}.  Due to the lack of the breakage specifically caused by mixed trackers, we did not train a separate \bd{} for mixed trackers and use the same breakage detector we used for non-mixed trackers.

\section{Evaluation}

We demonstrate the effectiveness of \name{}'s breakage reconstruction method and its ability to handle both non-mixed and mixed trackers.

\subsection{Breakage Reconstructability} \label{eval:breakage}

We aim to answer the following research questions
\begin{enumerate}[nolistsep,leftmargin=3.5em,label=Q\arabic*:]
    \item Can \name{}'s method of flipping exception rules reconstruct breakages?  (\S \ref{eval:breakge:q1})
    \item How does \name{}'s method of breakage reconstruction compare to previous works? (\S \ref{eval:breakge:q2})
\end{enumerate}

\begin{table}[t]
    \centering
    \caption{Result of of breakage reconstructability evaluation.}
    \rowcolors{2}{white}{gray!15}
    \begin{tabular}{p{5cm}p{1cm}p{1cm}}
             \rowcolor{gray!50}   
Description             & Exception rules & Commit message \\
        Reconstructable\&properly reconstructed  & 22 & 15 \\
        Reconstructable\&incorrectly reconstructed  & 0 & 6 \\
        Non-reconstructable & 18 & 19\\
    \end{tabular}
    \vspace{-3mm}
    \label{tab:breakage_evaluation_accuracy}
\end{table}

\subsubsection{Q1: Exception Rule-based Breakage Reconstructability} ~\label{eval:breakge:q1}
We conducted an experiment to determine during \name{}'s training set construction: 1) do the reconstruction heuristics properly reconstruct breakage? 2) What percentage of breakages are non-reconstructable by \name{}'s training process?

\subhead{Evaluation Dataset} 
We reconstruct breakages based on user reports containing ground truth descriptions. Our dataset was constructed by: 1) Finding all exception rules in EasyList, uBlock Origin, and AdGuard repositories as of June 1, 2024, and 2) Filtering out rules not referenced by user reports created between January 1 and May 31, 2024. We randomly sampled 40 rules from 142 exception rules to form our evaluation dataset.

Like previous work~\cite{blocked_or_borken}, we consider breakage to be ``reconstructable'' if the flipped commit blocks at least one resource. However, just because a resource is blocked, this does not mean that the page is necessarily broken, as the page may continue to work even if some resource is blocked. Thus, we also check whether breakage is properly reconstructed.
Our results in Table \ref{tab:breakage_evaluation_accuracy} show that: 1) All breakages that are reconstructable by \name{} are also properly reconstructed as they matched user report descriptions, and 2) \name{} found 45\% of the breakages to be non-reconstructable.

\subsubsection{Q2:Comparison to Commit Messages}~\label{eval:breakge:q2} We compare our method of breakage reconstruction with that of \cite{blocked_or_borken} (referred to as BoB thereafter), which uses commit messages as opposed to exception rules.
BoB begins by collecting breakage-fixing commits as determined by commit messages, and attempts to ``flip'' such commit into changes that cause breakage.  
\name{} is not directly comparable with BoB as they operate on different subjects: \name{} works with \textit{current} exception rules in filter lists, while BoB works with \textit{historic} commit messages. 
Still, we implemented BoB for comparison to answer the same questions in the previous evaluation: 1) when BoB indicates a breakage is reconstructable, does it properly reconstruct it? and 2) what percentage of breakages are non-reconstructable?

\subhead{Evaluation Dataset} We used the same method to collect breakage-fixing commits as described in \cite{blocked_or_borken}, filtering for commits referenced by user reports in EasyList, uBlock Origin, and AdGuard repositories between July 1, 2023, and May 31, 2024. 
This filtering was necessary for verification, as user reports provide the ground truth breakage descriptions. 
We randomly sampled 40 commits from 41 for our evaluation dataset. 

Our results in the third column of Table \ref{tab:breakage_evaluation_accuracy} show 6 cases where BoB indicated that a breakage occurred (i.e., was successfully reconstructed) but our manual investigation did not observe the same symptom as that described in the user reports. Such cases create noise in the training data. We manually investigated the causes of failure. We found that for 4 cases, the page changed between the breakage reconstruction time and breakage-reporting time (by comparing the current version with the closest version before the breakage-reporting time, using Wayback Machine~\cite{WaybackM71:online}). This means that, at breakage reconstruction time, the web pages no longer have any breakage and successfully blocking the breakage-fixing resource no longer reproduces the breakage. We were unable to confirm the remaining two cases. However, the root cause of our inability to properly reconstruct breakage is that commits cannot be removed from git repositories, and so they cannot take into account changes to the web page after the commit is made. The exception rule-based method does not suffer from this as rules can be updated---exception rules that do not trigger breakages tend to be actively removed by developers. 

During our evaluation, we found two general issues with BoB that, although they did not lead to incorrect reconstructions in our sample, they could still have potentially introduced noise:
\begin{enumerate*}
 \item Fixes requiring multiple commits: this occurs when a breakage is fixed with multiple commits rather than a single commit. This can happen when an initial breakage-fixing commit did not fully fix the breakage or caused other breakages, hence the need for subsequent commits. We observed 6 such cases in our sample. Multiple commits directly contradicts with BoB's assumption that a breakage-fixing commit contains all necessary changes to fix a breakage and flipping them reconstructs the breakage, leading to incorrect reconstruction.  

 \item Tangled commits: a tangled breakage-fixing commit includes both relevant and irrelevant changes for fixing the breakage. We observed 3 such cases in our sample. Flipping tangled commits can lead to noise in the reconstructed breakage due to the irrelevant changes. 
\end{enumerate*}

These issues arise because a commit is not always a good representation of breakage-fixing changes. A single commit may contain more changes than necessary to fix a breakage (tangled commits) or insufficient changes (multiple commits). In comparison, exception rules represent the exact changes to fix a breakage, as developers use them to specify the exact resource necessary for fixing breakages, avoid these issues.

\subsection{Non-mixed Trackers Detection} 

\noindent In non-mixed tracker evaluation, we answer the following questions:  
\begin{enumerate}[nolistsep,leftmargin=3.5em,label=Q\arabic*:]
    \item Do the detectors have predictive power? (\S \ref{sec:eval:q1})
    \item As a base test, how does \name{}'s accuracy compare to manually constructed \el{} such as EasyList and EasyPrivacy? 
    Can it identify additional trackers unreported on the \el{}? 
    Can it identify breakage caused by filter lists? (\S \ref{sec:eval:q2})
    \item Does \name{} exceed the state-of-the-art non-mixed tracker identification work? (\S \ref{sec:eval:q3})
    \item Other than discovering new trackers, how does  \name{} benefit the content blocker community? (\S \ref{sec:eval:q4})
\end{enumerate}

\subsubsection{Q1: Detector Accuracy} \label{sec:eval:q1}

We perform standard 5-fold cross-validation on the detectors to establish baseline classification accuracy on the training set. We report the mean F1 scores and standard deviations in Table \ref{tab:models_test_accuracy}. With mean F1 scores of 0.9591 and 0.9268 for the breakage and \td{}s, respectively, we conclude that the models are correctly trained and have predictive power.

\subsubsection{Q2: Comparison to Filter Lists}  \label{sec:eval:q2}

\begin{table}
    \centering
        \caption{Summary of the crawl used to evaluate \name{} on non-mixed tracker identification. D stands for \name{}, A stands for AdGraph.}
    \label{tab:crawl}
    \rowcolors{2}{gray!15}{white}
    
    \begin{tabular}{p{2.95cm}ccc}

   \rowcolor{gray!50} & Filter lists eval & \multicolumn{2}{c}{AdGraph eval } \\
   
   \rowcolor{gray!50} & (\S \ref{sec:eval:q2}) & \multicolumn{2}{c}{(\S \ref{sec:eval:q3})} \\
   
       Measurement period & Oct 2023 & \multicolumn{2}{c}{June-July 2024}\\
Crawl list & Alexa Top 1M & \multicolumn{2}{c}{Tranco List} \\

   \# pages  analyzed &  6,489   & D:2,645 & A:2,335 \\

       Size on disk (compressed)  & 3.2 TB & D:1.2GB & A:154MB\\
      
    Avg \# requests of a page & 142.13 & D:133.48 & A:114 \\ 
    \end{tabular}
\vspace{-3mm}
\end{table}

One problem with using \el{} as the ground truth for comparison is that they are imperfect: they may miss trackers and cause breakages. Thus, we first compute a tentative accuracy that does not account for cases where \name{} is correct and \el{} are wrong. 
We then perform a disagreement analysis where we manually inspect and assign a label to a sample of the disagreements between \name{} and \el{}. 
Using the disagreement analysis result, we can calculate an adjusted accuracy accounting for \el{}' inaccuracies. 
Finally, we perform an ablation analysis by running \name{}'s detectors alone. We show that the \bd{} is essential for \name{}'s overall accuracy. 
We note that since EasyList and EasyPrivacy are not capable of handling mixed trackers and thus will not block any mixed trackers. In this evaluation, we treat any identified mixed tracker as non-tracker so it is in line with \el{}' labels. We evaluate \name{}'s capability on mixed tracker identification in \S \ref{sec:eval:q4}.

\subhead{Dataset} We construct our evaluation dataset using requests observed while crawling 15K pages from each of the top, middle, and bottom 5K of the Alexa Top 1M List generated in May 2022, covering a variety of tracking methods. We label requests by invoking \name{} on all requests fetching JavaScript or containing request parameters as they may transmit tracking identifiers. In total, we analyzed 53,217 requests. The crawling statistics are summarized in Table \ref{tab:crawl}

\subhead{Tentative Accuracy} By comparing \name{}'s labels to \el{}' labels, we calculated an accuracy of 96.53\%. This is tentative accuracy as it may include cases where \name{} is correct and \el{} are wrong. There are 1,905 disagreements: 898 ``raw false positives'' and 1,007 ``raw false negatives''. We use the term `raw' as the results are pending manual investigation.

\subhead{Disagreement Analysis} \label{sec:eval:q2:disagreements} We manually label the disagreements. 
We find that \name{} is capable of identifying trackers missed by \el{} and instances where \el{} causes page breakage. Using the ratio of true positives and true negatives in the analyzed disagreements, we estimate \name{}'s adjusted accuracy to be 97.44\%.

Our goal is to assign a request to one of following labels below:
1) Breakage: the request serves functionality and removing it causes site breakage,
2) Stale request: the request serves functionality yet removing it does not cause immediate breakage,
3) Tracker: the request is a tracker, and removing it does not lead to breakage.  
4) Undecided: we cannot decide if the request is a tracker or part of the functionality.  
For mixed tracker, where the request is a tracker, but its removal breaks the page, due to the lack of labels in EasyPrivacy, we treat such cases as breakage. 

Due to the dataset's size, we cannot analyze all disagreements. We sample 40 cases from each type of disagreement, totaling 80 cases for manual labeling.

\subhead{Methodology} We describe our methodology for assigning labels. To determine if a request is a tracker, we gather information from online documentation, the request's initiator, URL, headers, and response. First, we search for any official documentation of the request. Tracking requests, especially third-party ones, often have official documentation (e.g., Twitter~\cite{Tracking97:online}, Adobe~\cite{AdobeAna95:online}). We label a request as a tracker if the documentation clearly states its tracking purpose.
For requests without official documentation, we look for discussions or consensus on similar requests and adopt the agreed-upon decision. For instance, Google's gen\_204~\cite{youtubet19:online} is generally recognized as a tracker. We refrain from labeling a request as a tracker if documentation or discussions indicate its involvement in A/B testing, integration, or optimizations, as we found in three cases.
Next, we examine the request's initiator. If the initiator is a known tracking script, we deem the request to be tracking. We also look for tracking-related keywords such as `track' and `analytics'. If a request has a response, we inspect the response content. Specifically, we read file-level header comments in JavaScript responses for references to documentation that describe the file's purpose (see Fig \ref{fig:header-comment} in Appendix \ref{appendix:sec:Breakage_examples} for an example). 
We also manually read unobfuscated or non-minimized code, determining the request as tracking if it involves fetching or sharing data with known tracking URLs. Finally, we check other resources hosted on the same domain using DuckDuckGo's Tracker Radar Wiki~\cite{TrackerR7:online} for potential clues. If the label remains undetermined after these steps, we categorize the request as undecided.

To determine if a request is legitimately required for functionality, we follow this procedure:
We search for documentation related to the request. Many legitimate requests are self-explanatory and well-documented. We label a request as legitimate if we find a documentation page that exactly describes it.
If documentation is unavailable, we perform a differential analysis to infer the request's purpose. We use two page instances: blocking the request in one and leaving it untouched in the other. We compare the behavior of the two instances, noting any differences that impact our ability to use the page. Specifically, we observe for visual breakages, such as missing content, iframes, images, and text. We also interact with the pages by scrolling, clicking links and buttons, resizing, hovering, and providing inputs to test input suggestions. We check if hyperlinks work but do not check the referenced page. If the request has a response, we examine it for hints on how the page may be impacted. For instance, if the response JavaScript interacts with a share button, we check if that share button on the page is broken.
We determine that a request is legitimately required for functionality if there is a clear association between the request and changes in page behavior. If the purpose of the request remains indeterminate after all procedures, we designate it as undecided.

\subhead{Raw False Positives} We first present the results of analyzing ``raw false positives'', where \name{} labels a request as a tracker but \el{} does not, summarized in the first four columns of Table \ref{tab:manual_analysis_result_non_mixed}. In these cases, \name{} was correct 55\% of the time. Through reporting newly discovered trackers to EasyPrivacy, we found that 30\% of these cases were new trackers not previously reported, and 25\% of the trackers could not be reproduced with \el{} enabled. We have two explanations: 1) probabilistic requests that do not always transmit, and 2) requests dependent on existing known trackers. All confirmed new trackers were reported to privacy developers, who added corresponding rules to \el{}. Non-reproducible trackers were not reported to avoid wasting the community's effort.

\begin{table}[t]
    \centering 
    \caption{Manual analysis results for a sample of ``raw false positives'' (\name{} labeled as trackers and filter lists labeled as non-trackers) and ``raw false negatives'' (\name{} labeled as non-trackers and filter lists labeled as trackers) . }
    \label{tab:manual_analysis_result_non_mixed}

    \rowcolors{2}{white}{gray!15}
    \begin{tabular}{cccc|ccc}
        \rowcolor{gray!50}     & \multicolumn{3}{c|}{False Positives} & \multicolumn{3}{c}{False Negatives} \\
        \rowcolor{gray!50}  & \# & \% & Estimated & \# & \% & Estimated \\
        Tracker  & 22 & 55 & 494 & 28 & 70 & 705\\ 
        Breakage  & 4 & 10 & 90 & 2 & 5 & 50 \\
        Stale requests  & 8 & 20 & 180 & 3 & 7.5 & 76\\
        Undecided  & 6 & 15 & 135 & 7 & 17.5 & 176\\
    \end{tabular}
    \vspace{-3mm}
\end{table}

\subhead{Raw False Negatives Analysis} Next, we analyze ``raw false negatives'', where \name{} labels a request as non-tracker but \el{} labels it as a tracker, summarized in the last three columns of Table \ref{tab:manual_analysis_result_non_mixed}. In these cases, 70\% were actual trackers, contributing to a high false negative rate. This is not surprising, as \el{} is used by billions of users, and breakages are promptly reported and fixed. We found that the false negatives were due to higher-than-expected predictions by the \bd{}, mistakenly labeling requests as legitimate when they aren't. Despite this, \name{}'s low false positive rate suggests that \name{} and \el{} find different sets of trackers and can complement each other. Additionally, due to the \bd{}, \name{} identified instances of \el{}-caused page breakages, detailed in \S \ref{sec:eval:q2:findings}.

\subhead{Adjusted Accuracy} Based on the numbers in Table \ref{tab:manual_analysis_result_non_mixed}, we calculate an upper bound for false positives as the sum of breakage, stale requests, and undecided cases, totaling 405 (90+180+135). The upper bound for false negatives, counting all known trackers, stale requests, and undecided cases, is 957 (705+76+176). We estimate \name{}'s accuracy rate to be 97.44\% (((22839+494)+(28473+50))/53217), referred to as the post-adjusted accuracy, and an F1 score of 0.9716. \name{}'s accuracy numbers are shown in Table \ref{tab:accuracies}, and we compare \name{}'s accuracy to the state-of-the-art in \S \ref{sec:eval:q3}.

\begin{table}[t]
    \centering
        \caption{\name{}'s accuracy numbers. }
    \label{tab:accuracies}

    \rowcolors{2}{white}{gray!15}
    \begin{tabular}{cc}
    \rowcolor{gray!50} Description & Accuracy (\%) \\
       Tentative accuracy  & 96.53 \\
       Adjusted accuracy  & 97.44 \\
       \Td{} only (ablation analysis \S \ref{sec:eval:ablation}) & 94.34 \\ 
    \end{tabular}
    \vspace{-3mm}
\end{table}

\subhead{\name{} Findings}  \label{sec:eval:q2:findings}
In this section, we detail \name{}'s findings. We identified 22 new trackers, with a total of 175 occurrences in our evaluation dataset, indicating that these trackers are not rare. 

We also detail two cases of EasyPrivacy-caused page breakages. One case involves \textit{ero-advertising.com}, where the script \textit{www.eroadvertising.com/js/controllers.js} is blocked by the EasyPrivacy rule \textit{/eroadvertising}. This overly generalized rule blocks the functional script \textit{controllers.js}, which loads the main body content on the site, resulting in a broken page missing the main body content, as shown in Fig~\ref{fig:fn2} in Appendix \ref{appendix:sec:Breakage_examples}.

Another example is on \textit{seznam.cz}, where the script \textit{ssp.seznam.cz/static/js/ssp.js?nocache=1} is blocked by the EasyPrivacy rule \textit{||ssp.seznam.cz\^}. This rule, designed to block trackers from \textit{ssp.seznam.cz}, overly generalizes and blocks \textit{ssp.js}, responsible for loading the cookie consent dialog. With EasyPrivacy on, the user does not see this dialog.

\subhead{Ablation Analysis} \label{sec:eval:ablation} In this analysis, we demonstrate that \name{} cannot achieve its accuracy without the \bd{}. We calculate \name{}'s accuracy using only its \td{}, tabulated in Table~\ref{tab:accuracies}. The results show that \name{} achieves its highest accuracy through a combination of the two detectors.

To confirm the \bd{}'s role in increasing accuracy, we manually analyzed the top 20 cases with the highest \bd{} predictions. Our analysis shows that 7 requests fetch general-purpose JavaScript libraries, 7 fetch JavaScript with specific functionality, such as push notifications and font-loading, and 6 are responsible for page content. We confirmed that blocking these requests leads to missing content, ranging from icons to sub-documents. We could not confirm the remaining two cases. This study shows that all analyzed requests are functional resources, indicating the \bd{}'s prediction power for detecting web page functionality resources. When used with the \td{}, the \bd{} can correct \td{}'s mispredictions on functional resources.

\subsubsection{Q3: Comparison to the State-of-the-art} \label{sec:eval:non_mixed_comparison} \label{sec:eval:q3}

\begin{table}[t]
    \centering
    \caption{Compare \name{} to AdGraph~\cite{iqbal2020adgraph} on 5K sites from Tranco List }
\rowcolors{2}{gray!15}{white}
    \begin{tabular}{ccc}
    \rowcolor{gray!50}
      Metrics   & AdGraph  & \name{} \\
    AuROC & 0.9669 & 0.9682 \\
    Accuracy (\%) & 93.51 & 93.85 \\
    Precision (\%) & 89.46 & 88.97\\ 
    Recall (\%) & 67.74 & 83.13 \\
    \end{tabular}
        \label{tab:adgraph_comparison}
    \vspace{-3mm}
\end{table}

We compared the accuracy of \name{} with AdGraph in classifying web requests as tracking or non-tracking. We selected AdGraph over other works ~\cite{siby2021webgraph, yang2022wtagraph}, because it was the only one we could execute successfully. 
Note that we used \name{}'s previously trained model (as mentioned in \S \ref{sec3:training_dataset}) for this evaluation.

We begin by describing our evaluation dataset. To enable a head-to-head comparison, we used both AdGraph and \name{} to crawl the top 5K sites from the Tranco List~\cite{pochat2018tranco} simultaneously, between June 22, 2024, and July 9, 2024. We tabulate the crawl information in Table \ref{tab:crawl}. Each tool uses its own browser to establish a web session with the web servers. The set of requests that each tool analyzed are different due to factors such as session-specific requests (e.g., URLs containing session IDs) --- we denote AdGraph's set as $R_A$ and \name{}'s as $R_D$. To form our evaluation dataset, we took the intersection of $R_A$ and $R_D$ to get a dataset that both tools analyzed excluding requests only analyzed by individual work. To prevent a few very commonly used trackers, such as Google Analytics, from dominating our results, we performed a de-duplication by request URL so that each request appears only once in our dataset. Our evaluation dataset contained a total of 18,122 requests. We first compare the accuracy of both tools against labels derived from EasyList and EasyPrivacy (referred to as filter lists thereafter).  Of the 18,122 requests, the filter lists label 15,233 as non-trackers, and the remaining 2,889 as trackers. We then tabulate the accuracy of \name{} and AdGraph at predicting the filter list labels in Table \ref{tab:adgraph_comparison}. We observed that AdGraph and \name{} achieved similar performance using filter lists as the ground truth. However, since filter lists are imperfect (e.g., \name{} found EasyPrivacy-caused breakages in \S \ref{sec:eval:q2}), in the next section, we further analyze instances where \name{} and AdGraph disagree.

\subhead{Disagreement Analysis} We conducted a manual analysis of the disagreements between AdGraph and \name{} to determine which method is more likely to make accurate tracker predictions. There are two types of disagreements: 242 requests where AdGraph labeled the request as a tracker while \name{} labeled it as a non-tracker, and 2,034 requests where AdGraph labeled the request as a non-tracker while \name{} labeled it as a tracker. We manually sampled 40 cases from each type of disagreement, totaling 80 requests. Each request was manually labeled as either a tracker, a non-tracker or undecided using the same methodology described previously in \S \ref{sec:eval:q2}.

Out of the 40 samples that were labeled by AdGraph as trackers, AdGraph was correct in 27 (67.5\%) of the cases and \name{} was correct in the remaining  13 (32.5\%) of the cases. Notably, we found two instances (from our sampled set, and five such requests in the whole set) of functional non-tracker requests being mis-labeled by AdGraph as trackers. Blocking such requests caused web page breakage. Both instances were requests on engadget.com that load images as part of the content from Yahoo's image resizing and optimization service. Blocking these requests broke the page as the images were absent; such requests share similarities with tracking requests syntactically. 
\name{}, on the other hand, was able to detect the breakage with its \bd{} --- the \bd{} returned a higher-than-threshold probability indicating that a breakage occurred when the request was blocked. As a system of two detectors, \name{} does not label such requests as trackers when the \bd{} detects breakage. Out of the 40 samples where \name{} labeled requests as trackers, 30 (75\%) requests are actual trackers that AdGraph failed to identify and 8 (20\%) requests were non-trackers. The remaining  2 (5\%) requests were undecided. We did not observe any cases where non-trackers caused breakage.

We note that \name{} labels far more requests as tracking and is also correct more often when it labels a request as tracking. By taking the rates at which \name{} is correct for the two disagreement types (13/40 and 30/40) and weighting them by the number of each disagreement (242 and 2034), we can expect \name{} to be correct in roughly 71\% of the cases when the tools disagree, demonstrating benefits of \name{}'s approach over AdGraph. In comparison, a similar analysis would reveal that AdGraph is only correct in 25\% of the cases where they disagree (we cannot estimate the remaining because of the 2 undecided cases).  
When \name{}'s accuracy compared to AdGraph is combined with the previous results from \S \ref{sec:eval:q2}, where \name{} was correct in 55.1\% (494/898) of the cases when \name{} labels a tracker and filter lists label it as a non-tracker, we believe \name{} is more likely to make the correct prediction compared to AdGraph and filter lists. Additionally, \name{}'s \bd{} was able to correctly identify two non-tracker functional requests where AdGraph misidentified them as trackers, causing page breakage.

\subsubsection{Q4: Benefit to the Community} \label{sec:eval:q4}

\subhead{Speed} On average, to analyze a single resource,  \name{} needs 363.64 seconds. This includes performing differential analysis (225.66s, 62.06\%), feature extraction (137.93s, 37.93\%) and invoking the detectors (0.05s, 0.01\%). The differential analysis step took the longest as this step is mostly CPU-bound: \name{} needs to launch two browser instances and load the page. \name{} is single-threaded, and thus, the computation time applies to our 2.4 GHz vCPU without any parallelism or GPU. We believe this setup is average, and the computation time is repeatable by others. In the evaluation, we parallelized 72 instances of \name{} on 72 cores and analyzed 53,217 requests  in slightly less than four days. This evaluation generated 23,737 rules, which translates to 
5,934 (23,737 / 4 days) rules per day. In comparison, \el{} insert 29.8 rules daily~\cite{snyder2018filters}.

\subhead{Cost} Based on current spot rate of USD \$0.00269/vCPU hour on Google Cloud at Las Vegas~\cite{PricingC80:online}, it takes USD \$0.02663 (\$0.00269*60*60 / 363.64s) to analyze a single candidate with \name{}. We estimate that it could cost USD \$ 1785.036 (0.02663*67,031) to analyze the whole dataset if ran on Google Cloud. This translates to an hourly cost of USD \$18.59. \name{} is cost-effective compared to human developers that manually identify trackers with an average hourly worker wage of \$48.32~\cite{TableB3A31:online}. Since \name{} is able to reproduce filter list labels quicker and at a lower cost, we believe it benefits the tracker detection community.

\subsection{Mixed Trackers Detection}
\label{sec:mixed_eval}

\noindent In mixed tracker evaluation, we aim to answer the following questions empirically: 
\begin{enumerate}[nolistsep,leftmargin=3.5em,label=Q\arabic*:]
    \item Can \name{}'s predictors effectively identify mixed request trackers?  (\S \ref{sec:mixed_eval:q1})
    \item Can \name{} stop mixed response trackers?  (\S \ref{sec:mixed_eval:q2})
    \item Can \name{} identify mixed request trackers in the wild?  (\S \ref{sec:mixed_eval:q3})
\end{enumerate}
We refrain from a direct comparison with previous work as we were unsuccessful in executing the previous work's code.

\subsubsection{Q1: Detector Accuracy} \label{sec:mixed_eval:q1}

We report the 5-fold cross-validation accuracy numbers in the last row of Table \ref{tab:models_test_accuracy}. 

\subsubsection{Q2: Stopping Mixed Response Trackers} \label{sec:mixed_eval:q2}
To answer whether \name{}'s method of blocking mixed request trackers can also stop requests with privacy sensitive information initiated by mixed response trackers, we first gathered mixed response trackers from our 15K page crawl described in \S \ref{sec:eval:q2}. Mixed response trackers are labeled by specific rules from UbO and AdGuard: redirect, replace, jsonprune, and hls. We identified 8,677 instances of mixed response trackers with demographics detailed in Table \ref{tab:mixed_response_tracker} in Appendix \ref{appendix:sec:Breakage_examples}.

We then collect requests initiated by mixed response trackers. To properly attribute outgoing requests to mixed response trackers, we first render the page containing mixed response trackers. Then, we rely on Chrome DevTool Protocol's \textit{requestWillBeSent} event. In the event handler, we check the request's initiator stack. A request is initiated by a mixed response tracker if the mixed response tracker appears anywhere in the initiating stack of that request.
In total, we collected 4,734 requests initiated by mixed response trackers. We again use \el{} to obtain a set of ground truth labels: 4,603 of 4,734 requests were labeled as trackers. This high percentage of trackers is expected because, as \el{} cannot block mixed response trackers directly, they instead block the requests that such trackers send out. Invoking \name{} on these requests, we achieved an accuracy of 95.39\%, indicating that \name{} can effectively stop mixed response trackers from sending privacy-sensitive information.

\subsubsection{Q3: Detecting Mixed Trackers in the Wild} \label{sec:mixed_eval:q3}

We evaluate \name{}'s ability to automatically identify mixed request trackers and assess their impact.

\subhead{Evaluation Dataset} To build our evaluation dataset, we starting from the complete 15K page crawl dataset used in the non-mixed tracker evaluation, and performs the following procedure:
\begin{enumerate}
    \item Filter out trackers by \el{}. We filter out all requests that \el{} label as trackers. 
    \item Filter out tracker-dependent requests. We filter out all requests that are not observable when \el{} are enabled: they are either probabilistic or dependent on \el{}-labeled requests.
    \item Expand into (request, request field) pairs. To prepare for \subr{} granularity analysis, for each request in the dataset, we expand it into (request, request field) pairs by parsing the URL parameters and cookie fields. Each parameter and each cookie will be a row in the dataset.   
    \item Filter out non-identifier request fields. At this step, we have a dataset of 251,038 (request, request field) pairs. Based on the average execution speed in \S \ref{sec:eval:q4}, it will take 2.89 years to exhaustively analyze all of them. However, we realize that not all request fields contain privacy-sensitive information: previous work has focused on identifier-like strings in network requests as they contain potentially privacy-sensitive information~\cite{PrivacyB7:online, 10.1145/2872427.2883028, PrivacyB7:online}. To compose a feasible dataset for evaluation, we adopt the same heuristic by filtering out all request fields that are non-identifier like. We leave the detail of this filtering step in Appendix. 
\end{enumerate}

Invoking \name{} on these fields, we obtained 7,133 positive labels and 11,302 negative labels. Lacking existing ground-truth labels for mixed trackers, and considering that advanced content blockers may block a broader range of information, we manually analyzed a subset by randomly sampling 40 cases from each prediction label, totaling 80 cases.

\subhead{Manual Analysis} Our goal is to assign each request field to one of the following labels: 
1) Breakage: if the field removal causes page breakage.
2) Stale field: if the field serves functionality but removing it has no impact on the page. 
3) Tracking field: if the field serves tracking, and field removal does not break the page. 
4) Undecided: if we cannot decide on the purpose of the request or the field.

\subhead{Methodology} We detail the procedure for assigning labels manually. 
While we follow the same method to assign request purposes as in \S \ref{sec:eval:q2:disagreements}, we detail how we assign field purposes and impact. 
Determining field purposes is challenging as they are server-designed and used; it is possible that a server names a field in a common way and uses it for a different purpose. We determine field purposes based on the following:
1) Documentation. We first try to search for any documentation on the field if possible. For query string parameters, we label a field as tracking if the documentation states that it is related to advertising, analytics or user identification. 
For cookies, we look up the cookie's purpose on Cookiepedia.
2) Field name and value. If the field has a common name, we look at how other parties use the same field if the field value format also matches (e.g., date, hash value). 
While this process is relatively easy for some query string parameters such as `hash' for hash value and `v' for version number, it is difficult to draw any conclusion for other query string parameters  like `token' or `sid'. We apply a conservative approach and assign a label of undecided if we have no concrete evidence to assign other labels. 
To assign field impact, we block the field using the `remoeveparam'  and `cookie' rule and use the same method described in \S \ref{sec:eval:q2:disagreements} to evaluate if the web page is broken.

\subhead{Positive Case Analysis} 
We present a summary of the analysis results in the first four columns of Table \ref{tab:merged:mixed}. Out of the positive cases that we analyzed, 65\% fields are potentially for tracking purposes. For instance, the `sessionId' parameter in Twitter's profile fetching and Tweet fetching API is potentially tracking. While we can never confirm whether it is used for tracking, we, through our manual analysis, found that this parameter is Twitter-set, has high entropy and removing it does not break Twitter's API on the page. We show one example on `myblogguest.com' in Fig \ref{fig:mix1} in Appendix \ref{appendix:sec:Breakage_examples}. None of the other parameters in the API are considered for tracking purposes. We detail additional findings at the end of this section. 
We found 10\% cases are page breakage. For instance, \textit{kyoto-ryokan.co.jp} sends a request to fetch TripAdvisor's certificate of excellence logo with a parameter `wtype=certificateOfExcellence'. While blocking this parameter changes the logo's appearance slightly, we deemed this as not tracking due to the nature of the logo. 
We also found 2.5\% to be parameters that have no immediate impact on the page but define an action that is executed conditionally at a later time dependent on user interactions. For instance, YouTube automatically tries to sign in to the user's Google account, and the request `https://accounts.google.com/ServiceLogin' contains a \textit{continuation} parameter that details the subsequent action after successful sign-in (continue=`https://youtube.com/signin'). Blocking this parameter has no immediate impact on the page but may break functionality if triggered by user interaction. Thus, for this type of interaction-dependent breakage, we introduce a new label:  potential breakage. 
We also found 15\% of the fields to be stale parameters. Blocking these fields has no impact on the server response.  We were unable to decide on the remaining 7.5\% of the fields.

\begin{table}[t]
    \centering 
    \caption{Manual analysis results of the positive and negative cases.}
    \label{tab:merged:mixed}

    \rowcolors{2}{gray!15}{white}
    \begin{tabular}{cccc|ccc}
        \rowcolor{gray!50} & \multicolumn{3}{c|}{Positive Cases} & \multicolumn{3}{c}{Negative Cases} \\
        \rowcolor{gray!50}  & \# & \% & Estimated & \# & \% & Estimated \\
        Tracking fields  & 26 & 65 & 4,636 & 2 & 5 & 565 \\ 
        Immediate breakage & 4 & 10 & 713 & 18 & 45 & 5,086 \\
        Potential breakage & 1 & 2.5 & 178 & 0 & 0 & 0 \\
        Stale fields & 6 & 15 & 1,070 & 14 & 35 & 3,956 \\
        Undecided & 3 & 7.5 & 535 & 6 & 15 & 1,695 \\
    \end{tabular}
    \vspace{-3mm}
\end{table}

\subhead{Negative Case Analysis} We present a summary of the analysis results in the last three columns of Tab \ref{tab:merged:mixed}.
We found 5\% cases are tracking fields that \name{} mistakenly predicted as functional. 45\% cases would cause breakage if blocked, 35\% are stale fields and 15\% cases are undecided. 

Using the estimated numbers, we can calculate a lower bound for \name{}'s accuracy in detecting mixed request fields to be 74.19\%((4,636+5,086+3,956)/18,435). 
We believe that our accuracy can be improved with a better training dataset as our current training dataset contains noise as discussed in \S \ref{sec:2}.

\subhead{\name{} Findings} While we believe there are more trackers to be discovered in our dataset, we measured the occurrence of confirmed trackers since our dataset consists of unique trackers. We observed a total of 83 occurrences in our evaluation dataset. We detail additional findings below. 
One case involves \textit{yandex.com}, which uses \textit{yandex.com/portal/set/any} to store user session data such as configuration data, considered legitimate site functionality. However, \name{} determined one query string parameter, \textit{szm=1:800x600:780x476}, to be tracking, as strings containing resolutions are commonly used for device fingerprinting. Removing this parameter had no impact on user experience across sessions in our manual analysis. Another case involves \textit{aaalife.com}, which loads \textit{\url{siteintercept.qualtrics.com...FeedbackButtonModule.js?Q_CLIENTVERSION=1.106.0&\_CLIENTTYPE=web&Q\_BRANDID=aaalife}}, a script rendering and supporting the feedback button, deemed legitimate page functionality. Upon closer inspection of its query string parameters, \name{} identified \textit{Q\_BRANDID} as a tracking parameter because it contains the source domain of the page. While such analytical information also appears in the Referer header, \name{} helps users who blocks this header ensuring such information is not leaked.

\section{Related Work}

\subsection{Tracker Identification}
We discuss proposals by the proposed feature source.

\subhead{Based on Network-level Information}
Several works hypothesize that trackers exhibit distinctive characteristics at the network level and thus extract features from network requests and responses. 
Bhagavatula et al.\cite{bhagavatula2014leveraging} focus on request query parameters, while Gugelmann et al.\cite{gugelmann2015automated} construct features from both requests (e.g., partiness) and responses (e.g., response size). Additionally, detecting cookie syncing has employed request and response features~\cite{papadopoulos2019cookie}.

A core insight used in many works, including \name{} for its mixed tracker evaluation, is that tracking relies on communicating identifier-like strings to servers. Yu et al.\cite{10.1145/2872427.2883028} emphasize the crucial role of such identifiers in tracking. The flow of unique values enables a server to distinguish clients\cite{PrivacyB7:online, 10.1145/2872427.2883028} or engage in cookie syncing~\cite{englehardt2016online}. Privacy Badger~\cite{PrivacyB7:online} employs the heuristic that a third-party domain is classified as a tracking domain if it sets unique identifier-like strings on more than three other domains.

\subhead{Based on JavaScript Features}
Another set of works hypothesizes that trackers exhibit different behavior (e.g., API invocations, DOM accesses) compared to functional JavaScript, extracting features based on static or dynamic JavaScript code features. Ikram et al.\cite{ikram2017towards} leverage syntactic and semantic JavaScript code features to detect trackers. Static analysis is vulnerable to obfuscation\cite{le2017towards}, making this approach less effective. On the other hand, Kaizer et al.\cite{kaizer2016towards} and DMTrackerDetector\cite{wu2016machine} rely on dynamic features, such as accessed properties, to build classifiers.

\subhead{Based on Rendering Graph} Recently, proposals have emerged that instrument the browser (e.g., Chromium~\cite{PageGrap4:online} and Firefox~\cite{siby2021webgraph}) to allow complete attribution of document modification, network resources, and JavaScript execution. These proposals model a web page's rendering process as a graph, from which works (\name{} included) extract features to predict trackers. All existing works train a single model for tracker detection. \name{} incorporates additional \bd{}, increasing overall accuracy while minimizing breakages.

\subsubsection{Mixed Trackers and Defenses}  
Amjad et al.~\cite{amjad2021trackersift, amjad2023blocking} conducted the first analysis of the prevalence of mixed trackers, revealing that 15\% of scripts are mixed trackers that cause breakage when blocked at the file granularity. One imprecision in Amjad's work is assuming that all requests not flagged by filter lists are functional requests. This is an overestimation as filter lists may not be complete, potentially inflating the number of identified mixed trackers.

SugarCoat~\cite{smith2021sugarcoat} is the first work addressing mixed trackers by attempting to automate the fixing of pre-identified mixed trackers. SugarCoat sanitizes mixed trackers by generating resource replacements where sensitive APIs are patched. While SugarCoat does not address the identification problem, it assumes that all privacy-sensitive APIs in a mixed tracker are used for tracking purposes. This assumption can lead to broken pages if the functional component of a mixed tracker also involves privacy-sensitive APIs, as patching such APIs can break the functional component. SugarCoat users must manually inspect to ensure no breakage.

ASTrack~\cite{castell2023astrack} operates on the observation that much third-party tracking code is embedded into first-party contexts, sharing code structures across first parties. By leveraging developer-identified tracking code patterns, ASTrack automatically identifies other tracking code finding similar patterns in presumably mixed scripts. However, it is unclear how ASTrack will work on newly constructed trackers, meaning it can have low recall and potentially miss new trackers. \name{} does not directly identify mixed response trackers but can identify privacy-sensitive information sent by mixed response trackers (\S \ref{sec:mixed_eval:q2}), stopping their tracking activities.

\subsection{Web Page Breakage Evaluation}
One reason for web page breakage is the use of content blockers who reply on developer-generated filter lists to remain effective~\cite{snyder2018filters}. Breakage can occur due to mistakes in any step (identification, fixing, and breakage evaluation) of the developer workflow. Additionally, web page admins may intentionally mix trackers with functional resources to evade detection~\cite{amjad2021trackersift}, and blocking mixed trackers can lead to breakages. Very few works have systematically studied the symptoms of breakage. Mathur et al.~\cite{mathur2018characterizing} conducted a survey on content blocker usage, evaluating why users adopt or avoid content blockers and how frequently they encounter blocker-induced breakage along with its symptoms. While users reported that content blockers rarely break sites, users might lack the technical background to identify certain types of breakage. Nisenoff et al.~\cite{nisenoff2023defining} systematically studied breakages caused by content blockers, analyzing user reports of broken pages and providing statistics on broken symptoms (e.g., missing content, non-responsiveness). We designed our \bd{} by encompassing their comprehensive set of symptoms.

\subhead{Automatic Breakage Evaluation}   Breakage evaluation in current tracker blocker implementations is predominantly manual. Content blockers like Adblock Plus and uBlock Origin depend on user reports of page breakages~\cite{AdblockP89:online}. Research proposals like AdGraph~\cite{iqbal2020adgraph} rely on manual evaluation, which is subjective and does not scale with automatic tracker detection. 

Research proposals attempt automatic breakage evaluation. Yu et al.~\cite{10.1145/2872427.2883028} propose using the page reload rate to detect web page breakage, assuming users will reload a web page if they experience a broken site. This assumption relies on the user to notice and react to breakages, which can be demanding. Moreover, a page reload may occur for reasons not related to breakage (e.g., to keep a live session).
PriVaricator~\cite{nikiforakis2015privaricator} and ASTrack~\cite{castell2023astrack} use changes in web page appearance to determine breakage. A page is potentially broken if its appearance differs from a non-broken page. This approach can be inaccurate as 1) there exist breakages without visual change (e.g., non-responsiveness), 2) this method suffers from visual non-determinism (\S \ref{sec:design:fo:q2}) where page appearance changes without breakage. 
AutoFR~\cite{le2022autofr} uses changes in the number of images and text before and after a page change to determine breakage. However, its heuristic is oversimplified and incomplete as it does not account for other ways a user can perceive page breakages (e.g., non-responsiveness).
Blocked or Broken~\cite{blocked_or_borken} attempts to automatically predict broken pages based on features extracted from PageGraphs~\cite{PageGrap4:online}. While its goal is similar to \name{}'s \bd{} (one of \name{}'s components), the approaches differ in data collection and the used features . Blocked or Broken collects broken page samples through commit messages, which are less relevant (as the commit may be outdated) and less standardized (i.e., not all commit messages follow conventions) compared to \name{}'s exception rules. 
SINBAD~\cite{chehade2024sinbad} is another work for breakage prediction based on user report forums data. Similar to commit messages, forums data is also a record of breakage occurred in the past and can be outdated. 

\name{} designed its features by modeling externally-visible channels covering all breakage symptoms discussed in \cite{nisenoff2023defining}; it collects its training data from exception rules from current filter lists which are constantly checked by privacy develoeprs. Consequently, \name{}'s \bd{} achieves higher test accuracy (shown in Table \ref{tab:models_test_accuracy}) than previous works.

\section{Limitations}

First, the \bd{} currently only measures breakage that occurs immediately on the client side. This means that it is unable to detect 1) breakages that do not occur immediately (e.g.,  after user interactions) or 2) breakages that only involve server-side symptoms. 
To be able to model non-immediate breakages, \name{} will need precise instructions as to how the breakages can be triggered (e.g., interactions to perform). While some issue reports contain such instructions, and tools such as large language models can help extract such instructions, we leave it as future work. 
To be able to model breakages with server-side symptoms, server cooperation is needed, which is challenging as the server is the party that tracks. Second, \name{} only support stateless analysis (e.g., no support for authenticated user sessions) because the modified requests it sends out, during mixed tracker analysis, may affect server states leading to server-side breakages. 
Third, \name{} currently does not detect dependencies among requests/request fields. It is possible that blocking multiple requests/ request fields at the same time lead to an opportunity of tracker blocking, however, it is expensive (exponential to the number of request/request fields) to enumerate all such possible combinations and we leave it as a future work.
Last, a server may detect and react to the modified requests that \name{} sends out during mixed tracker detection. The reaction may include, and is not limited to, sending fake responses or refusing to provide service, preventing \name{} from working.

\section{Conclusion}
In this paper, we introduced \name{}, a framework designed to identify both non-mixed and mixed trackers. \name{} has two novel mechanisms: 1) the addition of a \bd{} into tracker detection, which enables \name{} to detect web page breakage due to misidentification of non-tracker as tracker and blocking mixed trackers. 2) the use of differential features which enables \name{} to identify tracking components contained in mixed trackers. 

Our results demonstrate \name{}'s accuracy of 97.44\%  in detecting non-mixed trackers when compared to labels provided by \el{} across 15K pages. Moreover, we uncovered 22 unreported trackers and identified 2 cases of page breakage due to \el{}. In mixed tracker evaluation, \name{} achieves an accuracy of 74.19\%. Our manual analysis led to the discovery of 26 mixed request trackers. By promptly reporting our newly identified trackers and breakages to developers, we anticipate that \name{} will prove valuable to the community.

\section*{Acknowledgments}

We thank the anonymous reviewers for their feedback, and the artifact evaluation committee members for feedback on our artifacts. We also thank Bright Data for providing us with access to their proxies, without which this work would not be possible. Funding for this work was provided in part by ONR Contract N00014-17-1-2889, a contract with Telus, and NSERC Alliance Grant ALLRP 586310-23. He Shuang was supported by Ontario Graduate (OGS) and Bell Canada Scholarships. David Lie is supported by a Tier 1 Canada Research Chair.

\bibliographystyle{IEEEtran}
\bibliography{main}

\appendices

\section{Identifier-like Field Filtering}

\name{} uses entropy to select request fields for evaluation. The main insight is similar to previous works: a request field is likely to be used for tracking if it contains high entropy. Tracking identifiers tend to have high entropy as they need to uniquely identifier a user.
One problem with using field entropy to find identifier is that an identifier may be assembled from multiple low entropy request fields. For instance, a tracking identifier may be transmitted as a single value in a single request field (e.g., uid=aX13nL) or it can be transmitted as query string parameters in several requests and only assembled on the server side. 

\name{} employs two methods for entropy calculation: 
1) per field: this method calculates the total entropy of a single request field by estimating the total number of combinations that the field can hold based on the character set size and the field length. 
2) per server: this method calculates the total entropy of query string parameters and cookies of all requests sent to a single server to determine if the server can potentially assemble a high-entropy identifier. 

For mixed tracker evaluation, \name{} uses two configurable thresholds, 1 billion for per field and 1 trillion for per server, for the entropy calculation and keeps all fields that identified by either methods. The threshold setting poses a trade-off between performance and precision.  
A low threshold produces a larger set of request fields for evaluation with more irrelevant ones, and thus \name{} runs slower, while a high threshold produces a smaller set of more relevant request fields for faster evaluation but it is at the risk of missing mixed trackers. We picked our thresholds based on the identifiers observed in current non-mixed trackers.

\begin{table}[t]
    \centering
    \caption{Compare \name{}'s entropy filtering methods to the trackers identified by EasyList and EasyPrivacy. Agreements refer to the number of requests where \name{} and filter lists agree. }
\label{tab:tracker_searcher}
    \begin{tabular}{cccc}
       \rowcolor{gray!50} Method & Agreements & \name{} Only & Filter List Only \\
        Per field & 2812 (60.36) & 1526 (32.76) & 320 (6.87)\\
       \rowcolor{gray!15} Per server & 2194 (47.1) & 2370 (50.88) & 94 (2.02)\\
    \end{tabular}
    \vspace{-1mm}
\end{table}

To show the effectiveness of our heuristics and the thresholds, we compare the number of requests with identifiers that \name{} flags on the top 100 sites of Alexa Top 1M List generated in May 2022, to those identified by human-constructed filter lists such as EasyList and EasyPrivacy. If our methods can identify a close super-set of all trackers, then it is effective in finding tracking identifiers yet saving analysis on fields that are unlikely to be trackers. 
We show the numbers in Table \ref{tab:tracker_searcher}. 
We see that both methods produce requests sets that contain the majority of \el{} identified trackers with minimum misses (6.87\% and 2.02\%) respectively. %

\section{\Bd{} Features} \label{appendix:sec:bd_features}
We list all differential features used in \name{}'s \bd{}. We show all appearance features in Table \ref{tab:appearance_features}, input handling features in Table \ref{tab:input_features}, request features in Table \ref{tab:request_features} and the remaining features in Table \ref{tab:rest_features}.

\newcounter{magicrownumbers}
\newcommand\rownumber{\stepcounter{magicrownumbers}\arabic{magicrownumbers}}

\begin{table}[t]
    \centering
    \scriptsize
    \caption{Appearance-related Features}
    \label{tab:appearance_features}
    \rowcolors{2}{white}{gray!15}
    \begin{tabular}{lp{1.9cm}|p{5.4cm}}
\rowcolor{gray!50} \# & Feature Name & Description \\
\rownumber & VIPS screenshot  & \cellcolor{white}\\
\rownumber & Cormer screenshot & \multirow{-2}{5.4cm} {Cosine similarity among the largest section of page screenshots as returned by VIPS~\cite{cai2003vips} and Cormer et al.~\cite{cormer2017towards}} \\
\rownumber & Main screenshot &  \cellcolor{gray!15} \\ 
\rownumber & Section screenshot & \multirow{-2}{5.5cm}{\cellcolor{gray!15} Cosine  similarity between the screenshot of the first main/section tag } \\

\rownumber & Feature vectors & Cosine similarity of the feature vectors obtained by passing the screenshots to EfficientNet~\cite{tan2019efficientnet}\\
\rownumber & Text& Cosine similarity of the bag of words of the document text \\
\rownumber & Readability text& Cosine similarity of the document text extracted from Trafilatura~\cite{barbaresi-2021-trafilatura}\\
\rownumber & Document style& Cosine similarity of the CSS classes \\
\rownumber & Structure similarity& Cosine similarity of the sequence of the HTML tags
 \\
\rownumber & HTML& A joint of style and structure similarity \\
\rownumber & Fonts & $\Delta$ between the loaded fonts\\
\rownumber & Color& $\Delta$ of the unique colors used  \\
\rownumber & Height & $\Delta$ document height \\
\rownumber & Canvas& \cellcolor{gray!15} \\ 
\rownumber & Audio & \cellcolor{gray!15} \\ 
\rownumber & Button& \cellcolor{gray!15}\\ 
\rownumber & Input& \cellcolor{gray!15}\\
\rownumber & Links& \cellcolor{gray!15}\\ 
\rownumber & Dom scripts& \cellcolor{gray!15} \\
\rownumber & Span&  \multirow{-7}{5.5cm}{ $\Delta$ of canvas/audio/button/input/link(\textit{a} tag)/script/span element } \\
\rownumber & Unloaded diff& $\Delta$ of the number of \textit{Window.before\_unload} event \\
\rownumber & CSS files& $\Delta$ of  the number of CSS files parsed \\
\rownumber & Videos small& \cellcolor{white}\\
\rownumber & Videos large&  \cellcolor{white}\\
\rownumber & Video sensitive size& \cellcolor{white}\\
\rownumber & Images small& \cellcolor{white}\\
\rownumber & Images large& \cellcolor{white}\\
\rownumber & Images sensitive size& \cellcolor{white}\\
\rownumber & Iframes small& \cellcolor{white}\\
\rownumber & Iframes large& \cellcolor{white}\\
\rownumber &  Iframes sensitive size&  \multirow{-9}{5.5cm}{The difference in the number of video/image/iframe tag elements of small/large and sensitive size. Small size means the total area of the element is less than 10px. Sensitive size means the width and height of the element matches are commonly used for ads. } \\
\rownumber & Ads iframes& The number of iframes that has no inner text \\ 
\rownumber & Ad highlighter& $\Delta$  of the number of ads as identified by a perceptual ad detector~\cite{storey2017perceptual}\\
\end{tabular}
\vspace{-1mm}
\end{table}

\begin{table}[t]
    \centering
    \scriptsize
    \rowcolors{2}{gray!15}{white}
    \caption{Input handling-related features. }
    \label{tab:input_features}

    \begin{tabular}{lp{1.9cm}|p{5.4cm}}
    \rowcolor{gray!50} \# & Feature Name & Description \\

\rownumber & Specific listeners&  \cellcolor{gray!15} \\
\rownumber &Generic listeners& \cellcolor{gray!15}\\
\rownumber &Sensitive listeners& \cellcolor{gray!15}\\
\rownumber &Critical listeners& \multirow{-4}{5.5cm}{$\Delta$ of the event listeners on specific/generic/sensitive/critical elements.}\\
\rownumber &Functionality related listeners& $\Delta$ of the event listener types that are commonly used to serve functionality. \\
\rownumber & Listeners &  $\Delta$ of all event listeners \\ 
\end{tabular}
\vspace{-1mm}
\end{table}

\begin{table}[t]
    \centering
    \scriptsize
    \caption{Request features. FP refers to first party, TP refers to third party. }
\label{tab:request_features}

    \rowcolors{2}{white}{gray!15}
    \begin{tabular}{lp{1.9cm}|p{5.4cm}}
    \rowcolor{gray!50} \# & Feature Name & Description \\

\rownumber & \# requests blocked& \\
\rownumber &\% requests blocked& \cellcolor{gray!15} \multirow{-2}{5.5cm}{The \# and the \% of blocked requests}\\
\rownumber &URL length& The length of the blocked requests\\
\rownumber &Total parameters& Total number of parameters in blocked requests \\
\rownumber &Ad dimensions& \# of requests that contain dimension-like string in its URL\\
\rownumber &\# semicolon& Total number of semicolons in the blocked requests \\
\rownumber &\# screen & \# of the word `screen' is in the blocked requests\\
\rownumber &\# FP in blocked requests& \# of times that the first party domain exists in the blocked requests\\
\rownumber &\# FP req blocked& \\
\rownumber &\# TP req blocked& \cellcolor{gray!15}\multirow{-2}{5.5cm}{\# requests that are  first/third party}\\
\rownumber &\# ad keywords& \# of ad keywords in the blocked requests\\
\rownumber &\# storage values out& Number of storage values (local and session storage and cookies) in the blocked requests \\
\rownumber &API static& The number of sensitive API calls that are usually used for fingerprinting in the blocked request responses\\
\rownumber & Eavl keyword & \# of times that the keyword `eval' occurred in the blocked request responses \\
\rownumber &Total response size& \\
\rownumber &Avg response size& \multirow{-2}{5.5cm}{Total/average sized of the blocked request responses}\\
\rownumber &Sensitive FP& \\
\rownumber &Sensitive TP& \cellcolor{gray!15}\multirow{-2}{5.5cm}{Number of first/third party sensitive requests. } \\
\end{tabular}
\vspace{-1mm}
\end{table}

\begin{table}[t]
    \centering
    \scriptsize
    \caption{Storage, temporal performance and device interface features.}
\label{tab:rest_features}

    \rowcolors{2}{gray!15}{white}
    \begin{tabular}{lp{1.9cm}|p{5.4cm}}
    \rowcolor{gray!50} \# & Feature Name & Description \\

\rownumber & Storage & \\
\rownumber & Session storage& \cellcolor{gray!15}\\
\rownumber & Cookies& \multirow{-3}{5.5cm}{$\Delta$ of local storage/session storage and cookies}\\
\rownumber & Load time& $\Delta$ of the page loading time in seconds \\
\rownumber &  Logs & $\Delta$ in console log\\
\rownumber &  Downloads & $\Delta$ in the \textit{downloadWillBegin} event \\
\end{tabular}
\vspace{-1mm}
\end{table}

\section{Examples} \label{appendix:sec:Breakage_examples}
\begin{figure}[t]
    \centering
    \frame{\includegraphics[width=\linewidth]{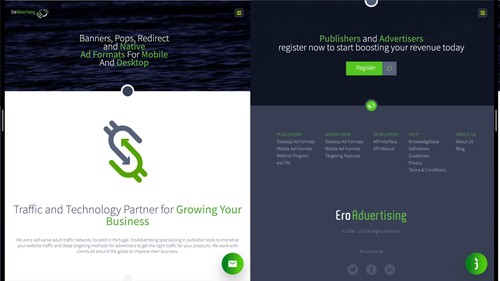}}
    \caption{\name{} discovered EasyPrivacy-caused site breakage on `ero-advertising.com'. }
    \label{fig:fn2}
    \vspace{-1mm}
\end{figure}

\begin{figure}[t]
    \centering
    \frame{\includegraphics[width=\linewidth]{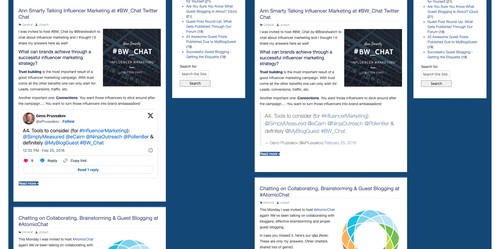}}
    \caption{\name{} discovered mixed tracker on `myblogguest.com'. }
    \label{fig:mix1}
    \vspace{-1mm}
\end{figure}

We show the following examples. 
\begin{enumerate}[nolistsep, leftmargin=*]
    \item An example of \el{}-caused page breakage is shown in Fig \ref{fig:fn2}. This example happens on `ero-advertising.com', where the script `www.eroadvertising.com/js/controllers.js' is erroneously blocked by EasyPrivacy on this domain causing the absence of main body content as shown on the right side of the figure. The functional page is shown on the left side. Although `ero-advertising.com' is an ads-serving domain, it is still a breakage as other users may want to visit the page (e.g., a customer of the company). Easylist and EasyPrivacy have historically fixed similar breakage~\cite{BlockBri6:online}.
    \item An example of \name{} discovered mixed tracker is shown in Fig \ref{fig:mix1}. We found the `sessionId' parameter in the request `https://platform.twitter.com/embed/Tweet.html'
    to be tracking on `myblogguest.com'. Removing this request breaks the page as the Tweet fails to display properly as shown on the right side of the figure. The functional page with the Tweet display is shown on the left.
\end{enumerate}

\begin{figure}[t]
    \centering 
    \frame{\includegraphics[width=0.9\linewidth, trim={0 26.8cm 33cm 0.2cm}, clip]{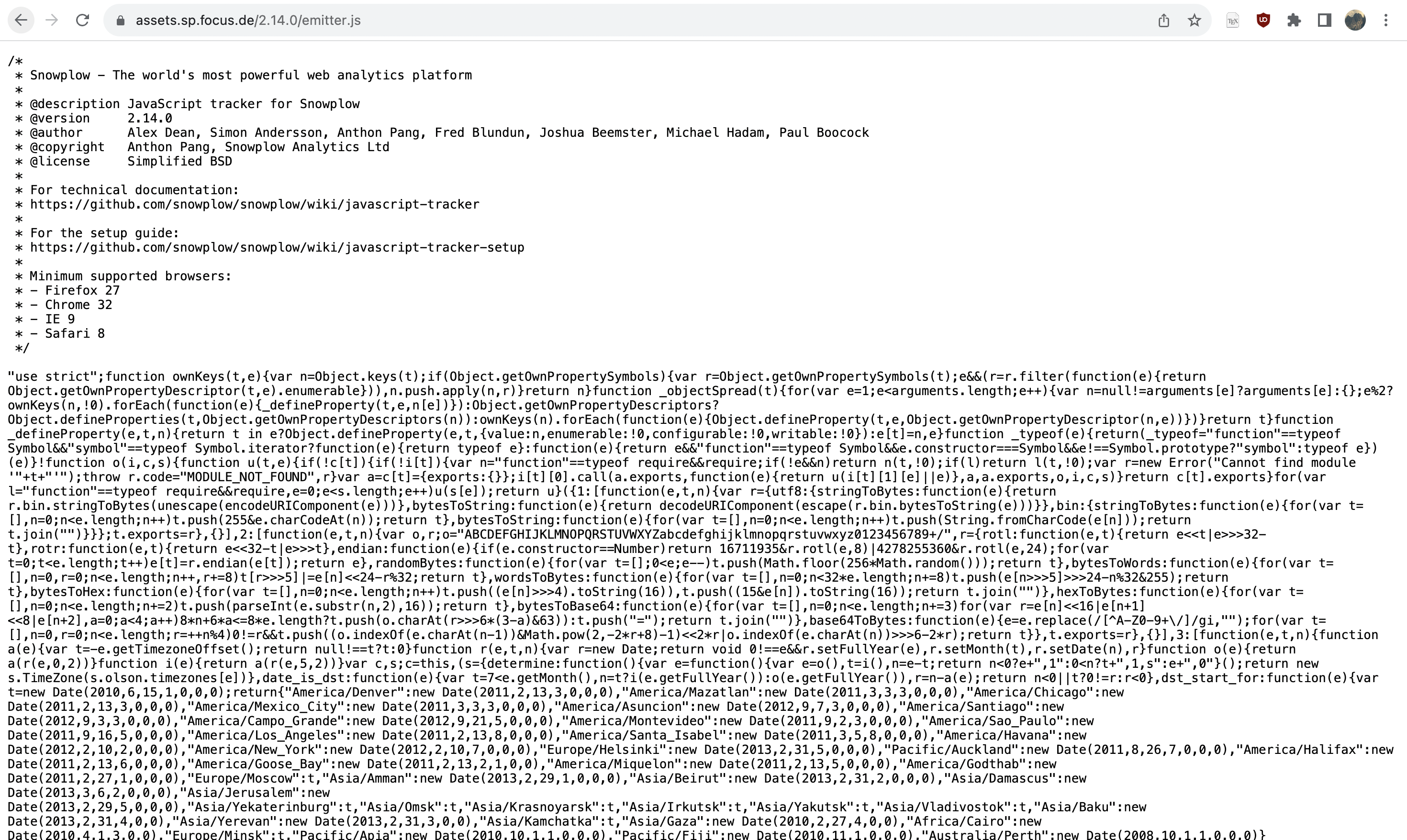}}
    \caption{A \name{} identified tracker on `focus.de' with file-level header comment that leads to external documentation. }
    \label{fig:header-comment}
\end{figure}

We also shown an example of trackers identified through external documentation in Fig \ref{fig:header-comment}.

We show the demographics of mixed response tracker from mixed response tracker evaluation (\S \ref{sec:mixed_eval:q2}) in Table \ref{tab:mixed_response_tracker}.

\begin{table}[!h]
    \centering
    \caption{Top 5 most frequent hostnames of mixed response trackers }
    \label{tab:mixed_response_tracker}

    \rowcolors{2}{white}{gray!15}
    \begin{tabular}{cc}
   \rowcolor{gray!50}     Mixed Response Tracker Hostname & Count \\
     \url{www.googletagmanager.com}    & 7556\\
     \url{securepubads.g.doubleclick.net}    & 663\\
     \url{www.googletagservices.com}    & 346\\
     \url{acdn.adnxs.com}    & 12\\
      \url{connect.mail.ru}   & 11\\
    \end{tabular}
\vspace{-1mm}
\end{table}

\section{Artifact Appendix} \label{appendix:artifact}
\subsection{Description \& Requirements}

\subsubsection{How to access}
The artifacts are publicly available on GitHub~\footnote{\url{https://github.com/dlgroupuoft/Duumviri-NDSS25}}. The \texttt{main} branch contains the latest version of the code. We utilize Docker to provide access to our working environment. The Docker image is available on Docker Hub and is also uploaded to Zenodo~\cite{Zenodo26:online}~\footnote{\url{https://zenodo.org/records/13621822}} as a tar file. Once downloaded, the tar file can be loaded into Docker with the following command: \texttt{docker load < ndss\_ae\_docker.tar.gz}

\subsubsection{Hardware requirements}
Our artifacts can be run on a commodity desktop machine with a x86-64 CPU. To ensure that all artifacts run correctly, a machine with at least 8 cores and 16 GB of RAM is recommended.

\subsubsection{Software requirements}
A recent Linux operating system and Docker are required. Our artifacts have been tested on Ubuntu 20.04 LTS (Focal Fossa) with Docker version 20.10.21.

\subsubsection{Benchmarks}
None.

\subsection{Artifact Installation \& Configuration}
Our docker image is available on Docker Hub~\footnote{https://hub.docker.com/r/8759s/ndss\_ae\_docker}. You can download our image with the following command:
\texttt{docker pull 8759s/ndss\_ae\_docker:latest}

You can run it with the following command:
\texttt{docker run -it --device /dev/fuse --privileged 8759s/ndss\_ae\_docker /bin/bash}

In this Appendix, we provide instructions for
running all experiments with this Docker image. Our GitHub repository contains more detailed instructions.

\subsection{Experiment Workflow}
Our artifacts contain four independent experiments. The
first experiment replicates \name{}'s ability in replicating EasyPrivacy and EasyList labels on non-mixed trackers. The second experiment loads and displays our manually sampled mixed tracker evaluation dataset. The third experiment detects non-mixed trackers on new sites. The fourth experiment detects mixed trackers on new sites. 

\subsection{Major Claims}

\begin{enumerate*}[label={(\bfseries C\arabic*):}]
\item \name{} achieves an accuracy of 96.53\% in replicating EasyList and EasyPrivacy labels on non-mixed tracker.
\item \name{} achieves an accuracy of 74.19\% in our manual mixed tracker analysis. 
\item \name{} can detect non-mixed trackers on new sites.
\item \name{} can detect mixed trackers on new sites.
\end{enumerate*}

\subsection{Evaluation}
This section includes all the operational steps. %

\subsubsection{Experiment (E1) - Claim (C1)}
This experiment 1) loads collected crawl data 2) determines the highest accuracy using EasyPrivacy and EasyList as ground truth by varying \bd{} and \td{} thresholds. The execution time is 11 minutes on our machine.  

\textit{[Execution]} Run ``\texttt{python3 code/eval.py replicate}''

\textit{[Results]}
The script will print the accuracy result. The result should be similar to the tentative accuracy 96.53\% reported in Section V.A.2.

\subsubsection{Experiment (E2) - Claim (C2)}
This experiment loads the manual analysis results discussed in \S \ref{sec:mixed_eval:q3} for inspection. This experiment loads df\_p and df\_n from positive.tsx and negative.tsx which contain the positive and negative prediction samples discussed in \S \ref{sec:mixed_eval:q3}.

\textit{[Preparation]} Go into \texttt{mixed\_tracker\_eval} folder. 

\textit{[Execution]} 
Run \texttt{python3 load.py} to load the data. 

\textit{[Results]} The script displays value counts statistics on our manual labels and spawns an interactive shell for further inspection.

\subsubsection{Experiment (E3) - Claim (C3)} 

This experiment detects non-mixed tracker on any site. This experiment will 1) load the page, 2) detect all requests to analyze, 3) analyze each request producing page deltas 4) creates features from page deltas and 5) invoke the trained models to produce output. The execution time depends on the number of requests to analyze. The analysis time for a request is roughly 6 minutes (as reported in \S \ref{sec:eval:q4}).

\textit{[Execution]} 
Run \texttt{./detect\_non\_mixed\_tracker.sh [URL]} to detect new trackers on any page specified by the URL.

\textit{[Results]} Requests that \name{} analyzed and the analysis result (i.e., non-mixed tracker or not).

\subsubsection{Experiment (E4) - Claim (C4)} 

This experiment detects mixed trackers on any site. This experiment will 1) load the page, 2) detect all fields to analyze, 3) analyze each field producing page deltas 4) create features from page deltas and 5) invoke the trained models to produce output. The execution time depends on the number of request fields to analyze. The analysis time for a request is roughly 6 minutes (as reported in \S \ref{sec:eval:q4}).

\textit{[Execution]} Execute the following commands to detect new trackers on any page specified by the URL.
\texttt{./detect\_mixed\_tracker.sh [URL]}

\textit{[Results]} Requests, request fields and value tuples that \name{} analyzed and the analysis result (i.e., mixed tracker or not)

\end{document}